\documentclass[showpacs,amsmath,amssymb,pra,aps,superscriptaddress,notitlepage,twocolumn,longbibliography]{revtex4-2}

\usepackage{amsmath,amssymb,amsthm,mathrsfs,amsfonts,dsfont}
\usepackage{qcircuit}
\usepackage[dvipdfmx]{graphicx}
\usepackage{subfigure, epsfig}
\usepackage{braket}
\usepackage{bm}
\usepackage{enumerate}
\usepackage{color}
\usepackage{comment}
\usepackage[normalem]{ulem}
\usepackage{here}
\usepackage{multirow}

\renewcommand{\sout}{\bgroup\markoverwith{\textcolor{red}{\rule[0.5ex]{2pt}{1pt}}}\ULon} 

\newcommand{\pdep}{p_{\rm dep}}
\newcommand{\pca}{p_{\rm cav}}
\newcommand{\pdel}{p_{\rm del}}
\newcommand{\psw}{p_{\rm sw}}
\newcommand{\pci}{p_{\rm cir}}
\newcommand{\ptn}{P_{{\rm tot},N}}
\newcommand{\ptf}{P_{{\rm tot},4}}

\newcommand{\len}{L_{\rm p}}

\newcommand{\nn}{\nonumber}
\newcommand{\om}{\omega}
\newcommand{\omc}{\omega_{\rm c}}
\newcommand{\ome}{\omega_{\rm 1e}}
\newcommand{\omg}{\omega_{01}}
\newcommand{\cdag}{c^{\dagger}}
\newcommand{\adag}{a^{\dagger}}
\newcommand{\cldag}{c_l^{\dagger}}
\newcommand{\crdag}{c_r^{\dagger}}
\newcommand{\ldag}{l^{\dagger}}
\newcommand{\rdag}{r^{\dagger}}
\newcommand{\lxdag}{l_{x}^{\dagger}}
\newcommand{\rxdag}{r_{x}^{\dagger}}
\newcommand{\kex}{\kappa_{\rm ex}}
\newcommand{\kin}{\kappa_{\rm in}}
\newcommand{\cin}{C_{\rm in}}
\newcommand{\ra}{\rangle}
\newcommand{\la}{\langle}
\newcommand{\rp}{\right}
\newcommand{\lp}{\left}

\newcommand{\fin}{f_{0,1}^{\rm in}}
\newcommand{\fout}{f_{0,1}^{\rm out}}
\newcommand{\raa}{\ra_{\rm a}}
\newcommand{\rap}{\ra_{\rm p}}
\newcommand{\raal}{\ra_{\rm a1a1}\la}

\begin{document}
\title{Fault-tolerant logical state construction based on cavity-QED network}

\author{Rui Asaoka}
\email{rui.asaoka@ntt.com}
\affiliation{NTT Computer and Data Science Laboratories, NTT Corporation, Musashino 180-8585, Japan}

\author{Yasunari Suzuki}
\email{yasunari.suzuki@ntt.com}
\thanks{R.A and Y.S contributed to this work equally}
\affiliation{NTT Computer and Data Science Laboratories, NTT Corporation, Musashino 180-8585, Japan}

\author{Yuuki Tokunaga}
\affiliation{NTT Computer and Data Science Laboratories, NTT Corporation, Musashino 180-8585, Japan}

\begin{abstract}
Exploring an efficient and scalable architecture of fault-tolerant quantum computing (FTQC) is vital for demonstrating useful quantum computing. Here, we propose and evaluate a scalable and practical architecture with a cavity-quantum-electrodynamics (CQED) network. Our architecture takes advantage of the stability of neutral atoms and the flexibility of a CQED network. We show a concrete framework for implementing surface codes and numerically analyze the logical error rate and threshold values beyond the simplified circuit-level noise model on several network structures. Although the requirement of CQED parameters is demanding given the current performance of experimental systems, we show that an error-decoding algorithm tailored to our proposed architecture, where the loss information of ancillary photons is utilized, greatly improves the error threshold. For example, the internal cooperativity, a good figure of merit of the cavity performance for quantum computing, required for FTQC is relaxed to 1/5 compared to the normal error-decoding for the surface code. Since our proposal and results can be extended to other LDPC codes straightforwardly, our approach will lead to achieve more reliable FTQC using CQED.
\end{abstract}
\maketitle

\section{Introduction}
\label{sec:intro}
Quantum computers can be used for solving a wide range of problems, including prime factoring~\cite{shor1997,gidney2021}, sampling from the solution of the linear systems~\cite{harrow2009}, and estimating the energy spectral of spin systems and molecules~\cite{babbush2018,kivlichan2020}.
To apply these algorithms to problems with a practical size, we need to integrate quantum error correction~(QEC) to achieve fault-tolerant quantum computing~(FTQC)~\cite{shor1995,shor1996,calderbank1996}. However, FTQC is still a daunting challenge because it is very demanding to manipulate an enormous number of qubits fast with a low error rate, adhering to the rules and structures required for QEC schemes.

With multiple promising approaches towards building FTQC in progress, approaches using neutral atoms have been outstanding in recent years~\cite{saffman2010,saffman2016,wang2016,levine2018,madjarov2020,browaeys2020,graham2022,evered2023,bluvstein2024}. It has several advantages: long lifetime of intrinsic states of atoms, high transportability with optical-tweezer arrays, and negligible correlated errors between atoms.
Instead, the neutral-atom processor has a drawback, weak interaction between qubits.
One strategy to compensate for the drawback is employing Rydberg states, which incur the interaction between selected qubits~\cite{saffman2010,saffman2016}. However, the scalability of this architecture is expected to be limited to up to a thousand qubits due to the trap capacity and challenging to connect multiple traps for further scale-up.
Another strategy is cavity quantum electrodynamics (CQED), which enables strong coupling between atoms and cavity field.
Although the scalability in a single cavity will be subject to similar limitations as Rydberg systems, {\it CQED network} including interconnects between cavities with trapped atoms has a promising potential to become a scalable quantum-computing platform~\cite{reiserer2022,covey2023,sunami2025}.
In recent years, CQED-network technology, such as nanofiber-cavity technology~\cite{kato2015,ruddell2020}, has been rapidly developing, where both strong coupling of neutral atoms to cavity field and seamless connection between cavities can be achieved. High designability in quantum processing with the itinerancy of photons opens the possibility to realize nearly all-to-all, i.e., high dimensional structure of qubit network required for building distributed quantum computing~\cite{yimsiriwattana2004,vanmeter2007,beals2013,vanmeter2016,caleffi2024} or low-overhead FTQC architecture~\cite{cohen2022,tremblay2022,yamasaki2024,bravyi2024,xu2024,pecorari2025}.

Quantum computing using CQED systems has been investigated for a long time in the field of qubit operations~\cite{pellizzari1995,zheng2000,duan2004,xiao2004,duan2005,koshino2010,reiserer2014,hacker2016,daiss2021} and communication between CQED-based quantum nodes~~\cite{reiserer2022,covey2023,sunami2025}. Although there are only a few, the fault-tolerance of a system consisting of connected multiple cavities has also been investigated, e.g., a visionary study by Goto and Ichimura~\cite{goto2010} estimating the fault-tolerant threshold regarding the C4/C6 code.
However, the previous studies have considered only abstract CQED-network structures, namely, have not explored the diverse design space with CQED-network. Furthermore, calculations in the previous studies remain within abstract model, e.g., a simplified noise model assuming the depolarizing channel following each gate, resulting that it is still unclear whether specific CQED network structures in realistic situation realize FTQC.
Thus, to set a promising course to build FTQC using a CQED network, it is an urgent issue to bridge the gap between experiments and theories.

\begin{figure*}[t]
    \centering
    \includegraphics[width=17.5cm]{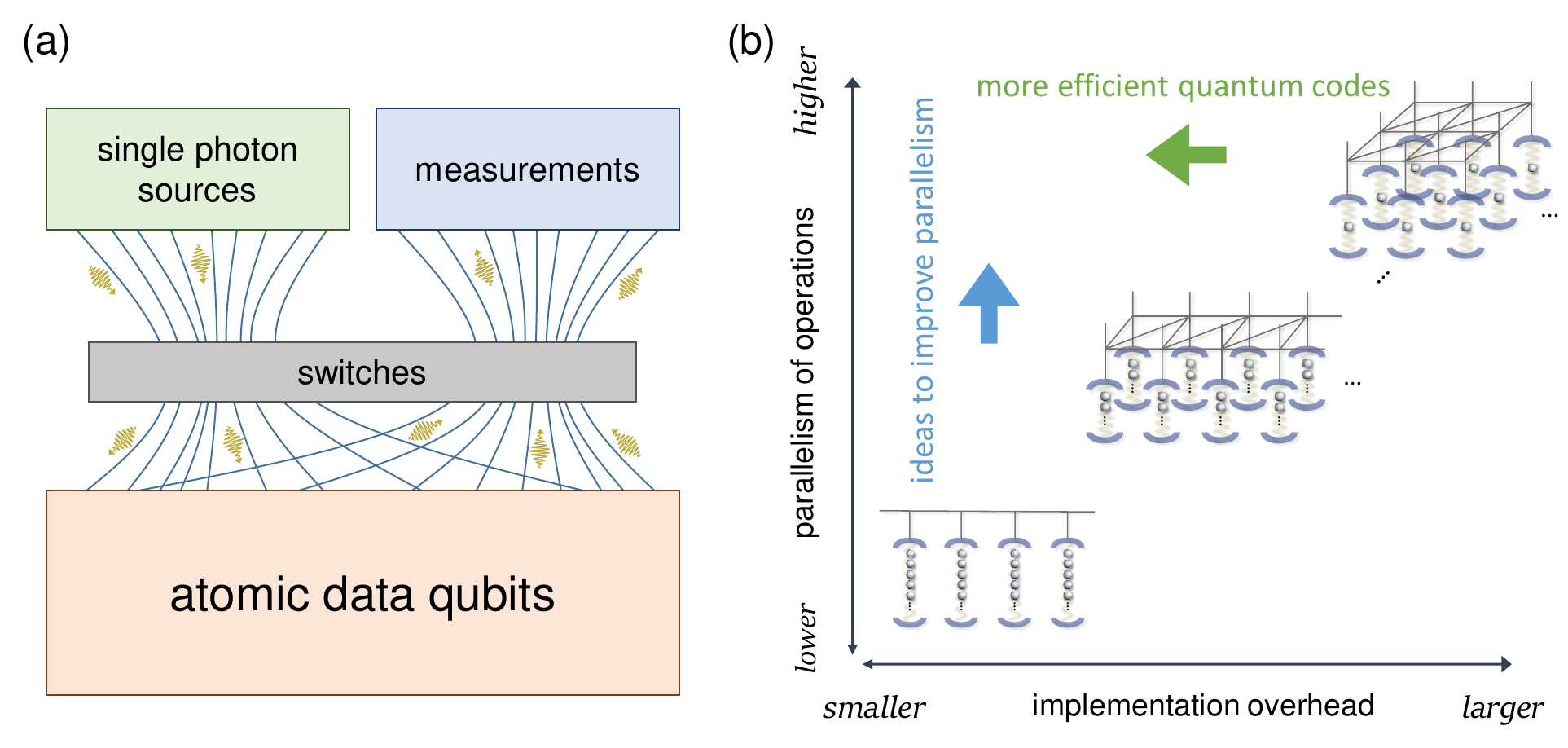}
    \caption{(a) Schematic of CQED network. (b) Trade-off relation of building a logical qubit based on CQED network.}
    \label{fig:ov}
\end{figure*}

In this paper, we propose a way to construct a logical qubit using several types of CQED-network structures and evaluate those fault-tolerance and scalability.
The basic structure of our proposal is shown in Fig.~\ref{fig:ov}(a). In this system, neutral atoms trapped in cavities are used as data qubits and photons as ancillary qubits. The ancllary photonic qubits can access the atomic qubits through optical switches, that is, by switching, you can change which atoms and in what order the photons access. This structure allows us to design a logical state flexibly only by controlling switchies, and the limitation of the arrangement of the atomic data qubits in real space is significantly relaxed thanks to the itinerancy of photons. Moreover, as explained in detail later, this structure also has an advantage that stabilizer measurements in this system can be {\it passively} performed without active control of atoms.
Based on this structure, we adopt the standard two-dimensional surface code to construct a logical qubit. As is well known, the surface code requires only the nearest-neighbor interaction between qubits; our architecture with itinerant photons has room for considering quantum codes which show higher performance, such as efficient low-density parity check codes~\cite{breuckmann2021,cohen2022,tremblay2022,bravyi2024,xu2024,pecorari2025}. Nevertheless, the surface code is superior in terms of the balance between the ease of its logical operations and the high threshold value and is an unavoidable stepping stone before exploring the most reasonable code.

We reveal the values of the CQED parameters, e.g., the coupling strength between atoms and cavity field, required for FTQC, where we treat with three types of dominant physical errors and analyze them on a stabilizer simulator without approximation. Since we also consider specific CQED network structures, our simulation can be directly compared to experiments in realistic situations.
The requirement of CQED parameters to achieve FTQC we derive in this work is demanding given the current performance of experimental systems. However, we show that an error-decoding algorithm tailored to our proposed architecture, where the loss information of ancillary photons is utilized, greatly improves the error threshold. For example, the internal cooperativity, a good figure of merit of the cavity performance for quantum computing, required for FTQC is relaxed to 1/5 compared to the normal error-decoding for the surface code. This indicates that there is room to significantly relax the conditions required for FTQC by error-tailored designing of QEC code, decoder, and total architecture. Our proposal and the results here can be extended to other LDPC codes, which may make use of the itinerancy of photons, and straightforwardly lead to exploring more reasonable code design.

This paper is organized as follows. We start by showing an overview and a basic approach for building a logical qubit with a CQED network in Sec.~\ref{sec:building}. Then, we introduce error sources of CQED network in Sec.~\ref{sec:err} and investigate their effect on the quantum circuit in Sec.~\ref{sec:erana}. In Sec.~\ref{sec:nume}, the error threshold values are numerically calculated based on stabilizer simulation, determining the CQED parameters required for achieving FTQC. Finally, Sec.~\ref{sec:discon} is devoted to summarizing our results and showing the perspective and possible future directions.
%
\section{Building logical qubits with CQED network}
\label{sec:building}
\subsection{Basic concept and overview}
\label{ssec:ov}
Basic structure for constructing logical qubits in Fig.~\ref{fig:ov}(a) includes atomic data qubits trapped in cavities, single photon sources, and measurement instruments, and they are connected through switches. Stabilizer measurements are performed by measuring ancillary (flying) photonic qubits, which gather parity information of data qubits via quantum gates with them. Ancillary photons are emitted from single photon sources and propagate along paths chosen using switches. Each path involves a group of data qubits that one would like to measure that parity.

There are a variety of ways to connect data qubits and the other components. In those CQED-network structures, the most reasonable implementation is determined considering several trade-off relations between indicators for building FTQC. In particular, the difference in CQED-network structures significantly influences the parallelism of operations and implementation overhead (Fig.~\ref{fig:ov}(b)). For example, when one atom is trapped per cavity, each connecting to a single photon source and a measurement instrument, parallelism of operations is highest, whereas many cavities are needed; that is, implementation overhead may be quite large. In contrast, when a cavity contains a number of atoms, implementation overhead is smaller, whereas parallelism of operations is low (it is generally difficult to make atomic qubits interact with flying photons independently and simultaneously as far as atoms in a cavity are resonant with the single cavity mode). In such a system, decoherence will have a greater negative impact on the fault-tolerance of a logical qubit.

In this paper, we propose logical qubit encoding based on CQED-network architectures and evaluate their error-correction capabilities and scalabilities.
Logical quantum information is encoded into the atomic qubits using the surface codes. We focus on the case of a single logical qubit for simplicity, that is, mainly discussing the fault-tolerance of the parity-check measurement process. Nevertheless, we would like to emphasize that our idea can be straightforwardly extended to the case of multiple logical qubits.

%
\subsection{Stabilizer measurement with an ancillary photon}
\label{ssec:stab}
\begin{figure}[t]
    \centering
    \includegraphics[width=8.5cm]{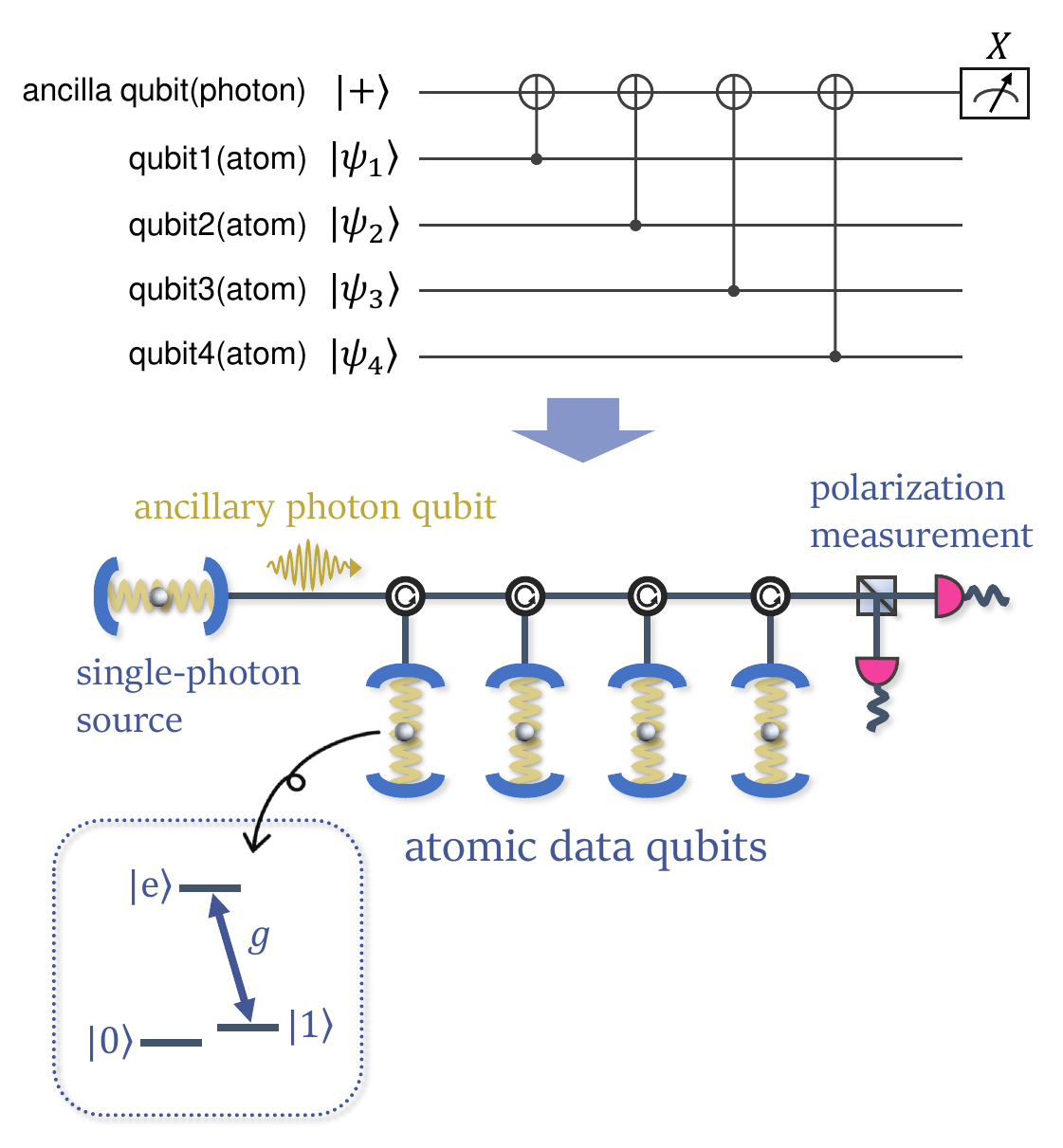}
    \caption{Stabilizer measurement with CQED network for constructing surface codes.}
    \label{fig:stab}
\end{figure}

In our protocol, error syndrome measurements rely upon non-demolition parity-check measurements through a sequence of entangling gates between atomic data qubits and an auxiliary single photon. Thanks to the strong interaction of atoms with optical modes via a cavity, we can perform the controlled-NOT (CNOT) gate between a flying photonic qubit and an atomic qubit. Therefore, with appropriate basis changes, we can perform a multi-qubit Pauli measurement with single-photon inputs, which is enough for performing stabilizer measurements.
Figure~\ref{fig:stab} shows a schematic picture of a stabilizer measurement for four atomic qubits, a typical one in the case of the surface code.
Ancillary photons pass through a CQED network, which indicates that stabilizer measurements in our architecture can be passively performed and that the limitation of the arrangement of the atomic data qubits in real space is significantly relaxed thanks to the itinerancy of photons.

Here we adopt the controlled-$Z$ (CZ) gate proposed by Duan and Kimble~\cite{duan2004} as a scheme for deterministic entanglement generation between photonic and atomic qubits. It utilizes {\it selective $\pi$-phase flip reflection} via the interaction between a three-level atom in a single-sided cavity as illustrated in Fig.~\ref{fig:stab} and a single-photon pulse. An atomic qubit is represented by the ground states, and a photon pulse carries a qubit represented by its polarization states. When the transition between $|1\ra$ and $|{\rm e}\ra$ states is resonant with only one polarization (e.g., circularly polarized state $|R\ra$), only the $|R\ra\otimes|1\ra$ state gains $\pi$ phase relative to other states due to vacuum Rabi splitting. If one would like to use linearly polarized light, which is unavailable for the $|1\ra$-$|{\rm e}\ra$ transition in general because of the selection rule, one can design the path of photons so that only one of the polarized states is incident to a cavity.
Thus, a sequence of the reflections of an ancillary photon pulse from single-sided cavities and the polarization measurement achieves the parity-check measurement. The cavities are connected with circulators and switches, which realize a reconfigurable CQED network, as detailed in Sec.~\ref{ssec:network}. 
%
\subsection{CQED-network structures}
\label{ssec:network}
\begin{figure}[t]
    \centering
    \includegraphics[width=8.5cm]{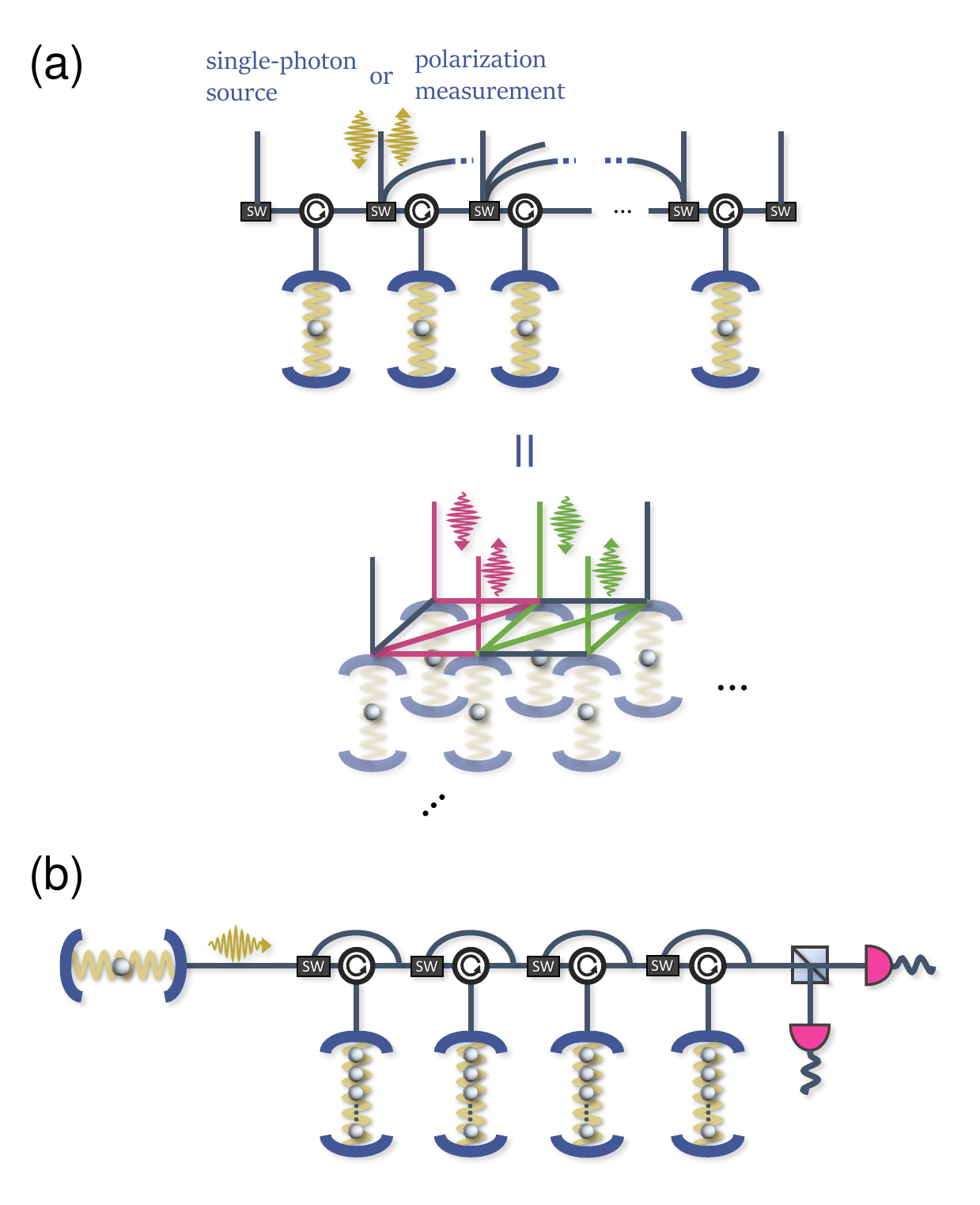}
    \caption{CQED-network structures for constructing surface codes. (a) Structure when the number of cavities is equal to the number of the atomic qubits, ${N=2d^2-2d+1}$. (b) Structure when only the minimum number of cavities, or four cavities, is available, where each cavity includes $(2d^2-2d+1)/4$ atomic qubits. One of the cavities is skipped with a corresponding switch when three-qubit Pauli strings on the edge of a logical qubit are measured.}
    \label{fig:nw}
\end{figure}

\begin{figure}[t]
    \centering
    \includegraphics[width=6cm]{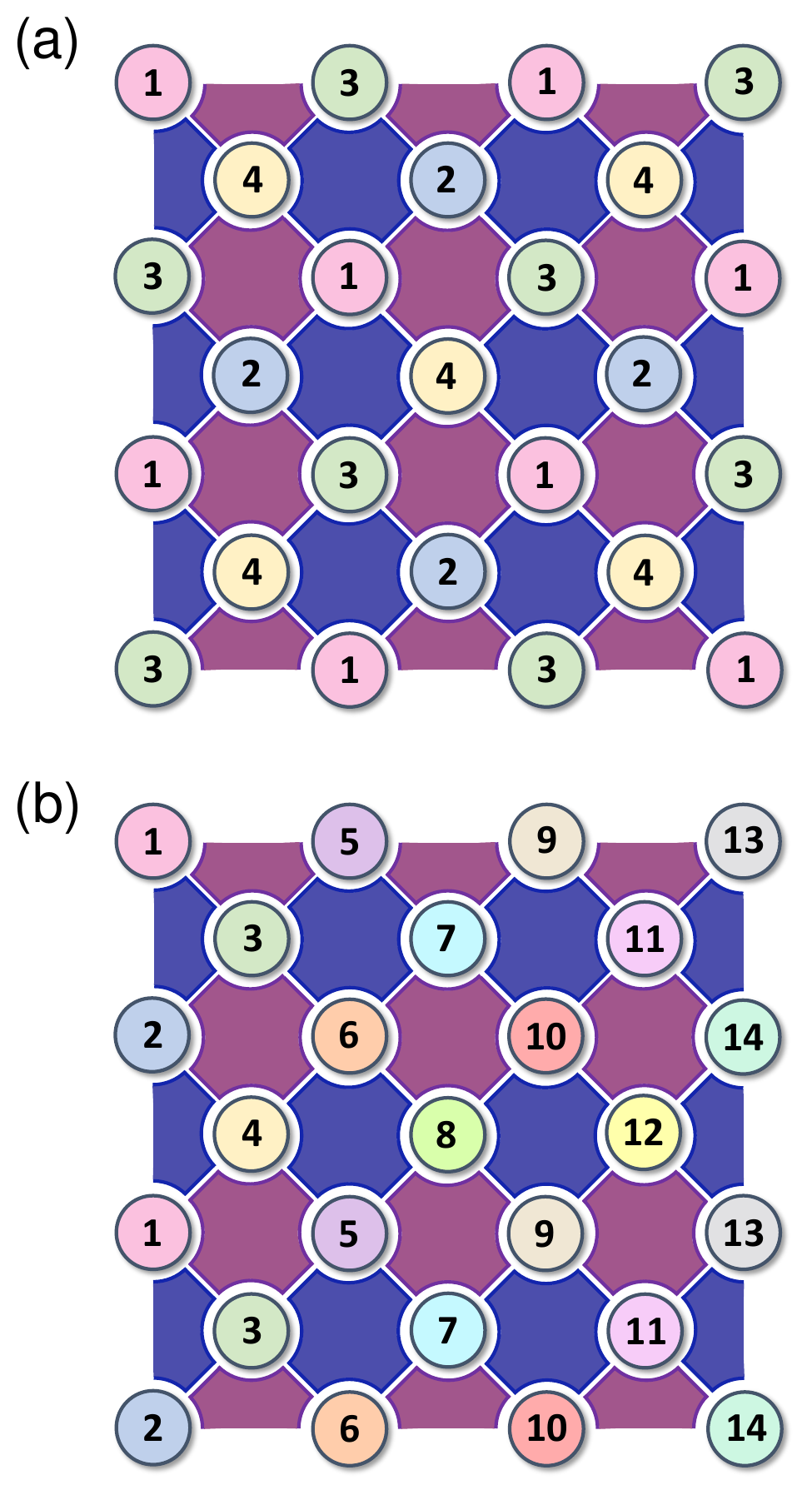}
    \caption{Assignment of cavity labels in the Q2D structure (a) with four cavities and (b) $O(d)$ cavities. Each circle corresponds to the data qubit in the surface codes, and purple and blue faces to Pauli-X and Z stabilizer measurements acting on the data qubits on its corners, respectively. By assigning qubits with the same color to the same cavity, all the stabilizer measurements act on at most one atom in each cavity.}
    \label{fig:lab}
\end{figure}

Here, we propose CQED-network structures for building a logical qubit with the surface code based on the fundamental structure in Fig.~\ref{fig:stab}. First, we consider two extreme CQED-network structures: $N$-cavity structure (Fig.~\ref{fig:nw}(a)) and $4$-cavity multi-atom structure (Fig.~\ref{fig:nw}(b)).
The former is a structure located at the top right in Fig.~\ref{fig:ov}(b), the most straightforward realization of the surface code; $N=2d^2-2d+1$ cavities ($d$ is the code distance), each including a single atomic qubit, are allocated in the two-dimensional (2D) grid array with the nearest neighbors connected (note that the cavities need not be arranged in the 2D grid in the real space). Switches can rearrange the path of photon pulses and even have the choice to connect polarization-measurement instruments or photon sources. This structure requires the same number of cavities as atomic qubits, but instead enables a highly parallel syndrome measurement if the cavities are connected in a two-dimensional grid where each node is connected to a single-photon source and a polarization-measurement instrument.
The latter has only four cavities, each including a 1D array of trapped atoms, which corresponds to the bottom left in Fig.~\ref{fig:ov}~(b). We can somehow choose which atoms to couple to the cavities, such as resonance shift depending on atomic position by gradient electric or magnetic field concerning a target atom, Stark shift by selective laser irradiation, or position shift of each atom by optical tweezers. This structure has the lowest parallelism of the syndrome measurements, or the largest syndrome-readout depth (equal to the number of syndromes $4(d-1)(2d-1)$), but instead requires the minimum number of cavities for constructing a logical qubit relying on the fundamental structure in Fig.~\ref{fig:stab}. Figure~\ref{fig:lab}(a) shows an example of the assignment of atomic qubits to four cavities so that the cavity labels do not overlap in any stabilizer measurements.

Thus, these structures are the extremes as for a trade-off relation between the experimental resource and the period of each syndrome measurement cycle or between the difficulty in implementing a logical qubit and the dephasing error due to a finite $T_2$.
Naturally, there are possible structures in between, where we would solve assignment problems of atoms to cavities. This problem is equivalent to {\it coloring problems}; the vertices on the square lattice are painted with as many different colors as the number of cavities, under the rule that the colors of the nearest and next-nearest neighbors are not the same.
An intermediate structure, for instance, is $O(d)$-cavity structure, where atoms are assigned to the cavities as shown in Fig.~\ref{fig:lab}(b).
On the other hand, the most reasonable implementation is strongly dependent on the circumstances of each experimental platform. Discussing such a specific problem in light of experimental developments is a future work.
%
%
\section{Errors in CQED-network architectures}
\label{sec:err}
%
\subsection{Error sources}
\label{ssec:source}
We expect three types of errors during a syndrome measurement cycle.
The first one is dephasing error due to a finite $T_2$. This forces the period of each syndrome measurement cycle to be sufficiently shorter than the lifetime. We note that $T_1$ can be neglected here because it is very long compared to other time scales characteristic of CQED systems.

The second is the infidelity of the atom-photon gate. The reduction in the fidelity of the CZ gate comes from unbalanced photon loss between the computational bases and distortion of reflected pulse shapes.
The former means that the photon loss probability depends on the states of atomic and photonic qubits. This causes the effective rotation of atomic qubits. However, it is known that the unbalanced photon loss can be canceled by designing cavity parameters appropriately~\cite{asaoka2021}, and we do not consider this error in this paper.
The latter is caused by frequency-dependent phase shift between the input (incident) and output (reflected) photon pulses~\cite{walls1994,asaoka2021}. 
In the frequency-dependent phase shift, the reduction in the gate fidelity is mainly caused by the first-order term regarding frequency or the delay in an output pulse~\cite{utsugi2025}.
Thus, in this paper, we consider that the infidelity of the atom-photon gate is due to the pulse delay (see Appendix.~\ref{app:delay} for more detail).  

The last dominant error source is photon loss through dissipative channels, namely undesirable scattering and absorption inside cavities, atomic spontaneous emission, transmission loss, and losses in detectors, circulators, and switches.
This photon loss error differs from the first two error sources in that this error can be detected, i.e., we can know that the photon is lost when the photodetectors do not click. We can improve the performance of the error correction by utilizing the photon-loss information, which will be discussed in Sec.~\ref{ssec:imp}.
%
\subsection{Simplified estimation of error accumulation}
\label{ssec:est}
To clarify the trade-off relations between the $N$- and $4$-cavity structures, including the photon loss and the pulse delay error, we first roughly estimate the error accumulation per data qubit in these structures. The error accumulation is here approximately defined as the sum of the error probabilities per atomic qubit during one error syndrome measurement cycle. A rigorous simulation of the logical error rate is performed in the next section.

In the $N$- and $4$-cavity structures, the error accumulation is given by
%
\begin{align}
\nn
\ptn &= \len\pdep \\   
\nn
&+ \frac{4(d-1)(2d-1)}{2d^2-2d+1}(\pca+\pdel+\psw) \\   
&+ \frac{2(d-1)(5d-2)}{2d^2-2d+1}\pci, 
\label{eq:ptn}  \\
\nn
\ptf &= 2d(d-1)\len\pdep \\     
\nn
&+ \frac{4(d-1)(2d-1)}{2d^2-2d+1}(\pca+\pdel+\pci) \\   
&+ \frac{8d(d-1)}{2d^2-2d+1}\psw  , 
\label{eq:ptf}
\end{align}
where we summarize the notations of the error probabilities and the physical quantities in Table.~\ref{tab:notat}. For example, in the $N-$cavity structure, the probabilities, $\pca$, $\pdel$, and $\psw$, are accumulated for the total number of cavities that the photons pass by. In this case, the number of the syndrome measurement of four- and three-qubit Pauli strings are $2(d-1)(d-2)$ and $4(d-1)$, and then the total number of cavities that the photons pass by is calculated as ${4\times2(d-1)(d-2) + 3\times4(d-1) = 4(d-1)(2d-1)}$. The number of times the photons passes through the circulators is one more than the number of times they pass by the cavities per syndrome measurement by the number of the circulators routing photons from single-photon source or to polarization measurement.
\begin{table*}
\caption{Notations of the error probabilities and the physical quantities.}
\begin{tabular}{p{34em}p{20em}}
\hline
\hfil dephasing rate per unit time& \hfil $\pdep$ \\
\hfil photon loss probability per the atom-photon CZ gate & \hfil $\pca$ \\
\hfil pulse distortion error probability per the atom-photon CZ gate & \hfil $\pdel$ \\
\hfil photon loss probability per switch & \hfil $\psw$ \\
\hfil photon loss probability per circulator & \hfil $\pci$ \\
\hfil pulse length & \hfil $\len$ \\
\hline
\end{tabular}
\label{tab:notat}
\end{table*}
Thus, the error accumulation does not increase with code distance $d$ in the $N$-cavity structure, whereas that increases in the 4-cavity structure because of the dephasing term. This indicates that the 4-cavity structure does not have a finite threshold value because the logical error rate for the surface code, simply written as $p_L\propto (p/p_{\rm th})^{\frac{d+1}{2}}$~\cite{fowler2012}, begins to increase with increasing $d$ beyond a certain code distance $d_{\rm c}$. This problem will be discussed in detail in Sec.~\ref{ssec:tole}.
To suppress the total dephasing error in the $4$-cavity structure, we should reduce the period of a syndrome-measurement cycle by making the pulse length $\len$ short. However, $\pdel$ increases for a shorter pulse. As a result, there is an optimal pulse length to minimize the total error.
%
\section{Error model analysis}
\label{sec:erana}
To calculate the logical error rate considering the physical noise discussed in Sec.~\ref{ssec:source}, we derive the noise map of the error sources and translate it to a form that can be treated in stabilizer simulation.
%
\subsection{Derivation of noise map}
\label{ssec:map}
Since the CPTP map of the dephasing error is obvious of the three error sources that we focus on in this paper, we now derive the error model that represents the photon loss and the pulse delay in the stabilizer measurements on the atomic data qubits.
First, we show an example of a photon loss event that would occur between the first and second CNOT gates (we model the photon loss at the first CNOT gate to be included in this case), and then results in other cases are shown briefly.

\begin{figure}[t]
    \centering
    \includegraphics[width=8.5cm]{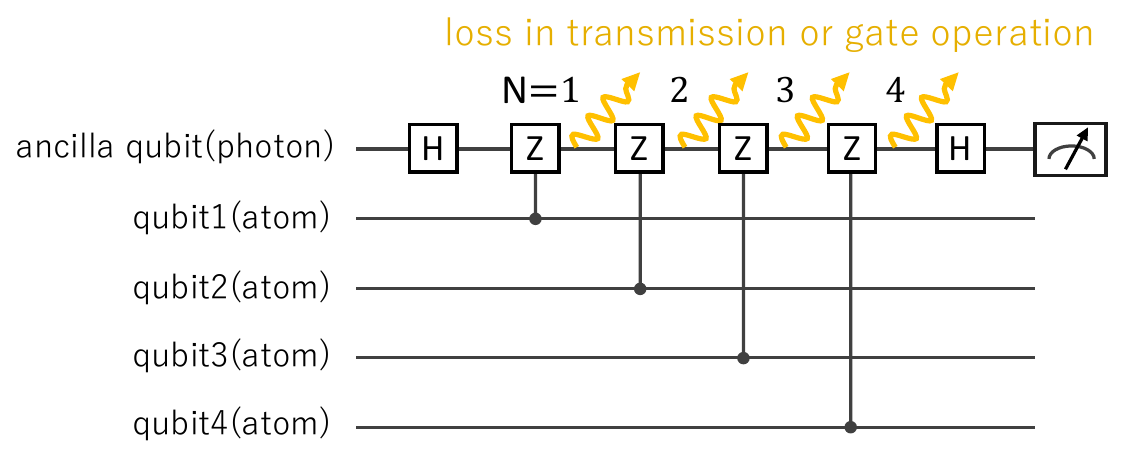}
    \caption{Quantum circuit diagram of $Z$ stabilizer measurement by CZ gates.}
    \label{fig:loss}
\end{figure}

The quantum circuit of the $Z$ stabilizer measurement in Fig.~\ref{fig:stab} can be translated to that in Fig.~\ref{fig:loss} with Hadamard gates and CZ gates.
Thus, the entire state of photonic and atomic qubits just before the first CZ gate is expressed as
\begin{align}
\nn
|\Psi_i^{(1)}\ra =
\frac{1}{\sqrt{2}}\int dxf_{i}(x)\lp(\lxdag+\rxdag\rp)|{\rm vac}\rap\otimes \\
\lp(\alpha_0|0\raa{}_1|\phi_0\ra_\perp + \alpha_1|1\raa{}_1|\phi_1\ra_\perp\rp),
\label{eq:pin}
\end{align}
where $|q\raa{}_1$ denotes the qubit state of the first atom and $|\phi_q\ra_\perp$ indicates the entire atomic state except for the first atom being in $|q\raa{}_1$. $\lxdag(\rxdag)$ is the creation operator of $L(R)$ polarization at a position $x$. The initial probability amplitudes are normalized as ${|\alpha_0|^2+|\alpha_1|^2=1}$ and ${\int dx|f_{i}(x)|^2=1}$.
It is known that the pulse delay error finally corresponds to the basis-dependent reduction in the probability amplitudes, that is, the imbalance reduction of $f_{i}(x)$, $\alpha_0$, and $\alpha_1$ regarding the photonic and atomic bases~\cite{utsugi2025}.
Therefore, if no photon loss has occurred up to that point, the output state of the atomic qubit 1 and the ancillary photon qubit after the first CZ gate is written as
\begin{align}
\nn
|\Psi^{(1)}\ra
= \frac{1}{\sqrt{2}}\int dx\lp[f_{l1}(x)\lxdag(\alpha_0|0\raa{}_1|\phi_0\ra_\perp \rp. \\
\nn
\lp. + \alpha_1|1\raa{}_1|\phi_1\ra_\perp)|{\rm vac}\rap \rp. \\
\nn
\lp. + \rxdag(f_{r0}(x)\alpha_0|0\raa{}_1|\phi_0\ra_\perp \rp. \\
\lp. - f_{r1}(x)\alpha_1|1\raa{}_1|\phi_1\ra_\perp)|{\rm vac}\rap \rp],
\label{eq:pout1}
\end{align}
where $f_{l(r)n}(x)$ is the probability amplitude of the $L(R)$-polarized photon pulse after the reflection from $n$th cavity (that is, the maximal number of $n$ is 4). However, for the $R$ polarization, the number of the reflections is counted only when the atom in $n$th cavity is in the state $|1\ra_{{\rm a}n}$. In Eq.~(\ref{eq:pout1}), we assume that the transition between $|1\ra$ and $|{\rm e}\ra$ is resonant only with the $R$ polarized photon, where the reduction in the probability amplitude of a $L$-polarized photon never depend on the atomic state.

Photon loss in the stabilizer measurement corresponds to taking the trace of the density operator on the position and the polarization bases. Thus, the density operator of the atomic qubit after the photon loss event ($N=1$ case in Fig.~\ref{fig:loss}) is given by
\begin{align}
\nn
\rho_{\rm a1} = &|\alpha_0|^2|0\raal 0| + |\alpha_1|^2|1\raal 1| \\
\nn
&+ \frac{\alpha_0\alpha_1^*}{2}\lp(1-\int dx f_{r0}(x)f_{r1}^*(x)\rp)|0\raal 1| \\
&+ \frac{\alpha_0^*\alpha_1}{2}\lp(1-\int dx f_{r0}^*(x)f_{r1}(x)\rp)|1\raal 0|,
\label{eq:den1a}
\end{align}
where we summarized the expression of $|q\raa{}_1|\cdots\raa{}_{1,q}$ as $|q\raa{}_1$. 
The integral terms in Eq.~(\ref{eq:den1a}) represent the discrepancy of the pulse delays on the atomic and photonic states (they are equal to 1 if there was no delay or all the delays are equal). By putting ${\int dx f_{r0}(x)f_{r1}^*(x)=\int dx f_{r0}^*(x)f_{r1}(x)=W_1}$ (note that $W_1$ can be assumed to be a real value in settings explained later), we finally derive the error model of the photon loss and the pulse delay represented by Pauli channels:
\begin{align}
\nn
\rho_{\rm a1} &= |\alpha_0|^2|0\raal 0| + |\alpha_1|^2|1\raal 1| \\
\nn
& \hspace{3mm}+ \frac{1-W_1}{2}(\alpha_0\alpha_1^*|0\raal 1|+\alpha_0^*\alpha_1|1\raal 0|) \\
\nn
&=
\begin{pmatrix}
|\alpha_0|^2 & \frac{1-W_1}{2}\alpha_0\alpha_1^* \\
\frac{1-W_1}{2}\alpha_0^*\alpha_1 & |\alpha_1|^2
\end{pmatrix}
\\
&= \frac{3-W_1}{4}\rho_{\rm a,ini} + \frac{1+W_1}{4}Z_1\rho_{\rm a,ini}Z_1,
\label{eq:den1b}
\end{align}
where $\rho_{\rm a,ini}$ denotes the initial density operator of the atomic qubit.

In the ${N=2}$ photon loss case, the output density operator of the atomic qubits 1 and 2 is similarly given by
\begin{align}
\nn
&\rho_{\rm a2} = \\
&
\begin{pmatrix}
|\alpha_{00}|^2 & \frac{1-W_1}{2}\alpha_{00}\alpha_{01}^* &
\frac{1-W_1}{2}\alpha_{00}\alpha_{10}^* & \frac{1+W_2}{2}\alpha_{00}\alpha_{11}^* \\
\frac{1-W_1}{2}\alpha_{01}\alpha_{00}^* & |\alpha_{01}|^2 &
\alpha_{01}\alpha_{10}^* & \frac{1-W_1}{2}\alpha_{01}\alpha_{11}^* \\
\frac{1-W_1}{2}\alpha_{10}\alpha_{00}^* & \alpha_{10}\alpha_{01}^* &
|\alpha_{10}|^2 & \frac{1-W_1}{2}\alpha_{10}\alpha_{11}^* \\
\frac{1+W_2}{2}\alpha_{11}\alpha_{00}^* & \frac{1-W_1}{2}\alpha_{11}\alpha_{01}^* &
\frac{1-W_1}{2}\alpha_{11}\alpha_{10}^* & |\alpha_{11}|^2
\end{pmatrix}.
\label{eq:den1b}
\end{align}
Here the initial state of the atom qubits is ${|\psi_i\raa=\alpha_{00}|00\raa+\alpha_{01}|01\raa+\alpha_{10}|10\raa+\alpha_{11}|11\raa}$, which is normalized as ${|\alpha_{00}|^2+|\alpha_{01}|^2+|\alpha_{10}|^2+|\alpha_{11}|^2=1}$, and $W_2$ is defined by ${W_2\equiv\int dx f_{r0}(x)f_{r2}^*(x)=\int dx f_{r0}^*(x)f_{r2}(x)}$. Unlike the case of ${N=1}$ photon loss, this density matrix cannot be described only by the Pauli channels. This is also the case with ${N=3}$ and 4.

For all $N$, the output density matrices can be expressed concisely by omitting the coefficients $\alpha_s$ to define the initial state. Letting $\{|A_k\ra\}$ be the states that include $k$ atoms being in the $|1\raa$, the matrix elements of $|A_k\ra\la A_{k+m}|$ except $\alpha_s$ are given by
\begin{align}
M_{k,k+m} = M_{k+m,k} = (1+(-1)^m W_m)/2 \hspace{3mm}(m=0,\cdots,4).
\label{eq:den}
\end{align}
Here, since ${\int dx f_{ra}(x)f_{rb}^*(x)=\int dx f_{rc}(x)f_{rd}^*(x)}$ when ${a-b=c-d}$, we define $W_m$ by
\begin{align}
\nonumber
&W_0=1,\\
&W_{m\neq 0}=\int dx f_{rm_1}(x)f_{rm_2}^*(x)=\int dx f_{rm_1}^*(x)f_{m_2}(x),
\label{eq:wm}
\end{align}
where $m_1$ and $m_2$ satisfy ${m_2-m_1=m}$.
\begin{table*}
\caption{Elements of the density matrix depending on the results in the polarization measurement.}
\begin{tabular}{p{15em}p{12em}p{12em}}
\hline \hline
\hfil & \hfil $L$-polarization & \hfil $R$-polarization \\
\hline 
\hfil $|A_0\ra\la A_0|, |A_4\ra\la A_4|$ & \hfil $\frac{1+W_2}{2}$ & \hfil $\frac{1-W_2}{2}$ \\
\hfil $|A_1\ra\la A_1|, |A_3\ra\la A_3|$ & \hfil $\frac{1-W_1}{2}$ & \hfil $\frac{1+W_1}{2}$ \\
\hfil $|A_2\ra\la A_2|$ & \hfil 1 & \hfil 0 \\
\hfil $|A_0\ra\la A_1|, |A_3\ra\la A_4|$ & \hfil $\frac{1-2W_1+W_2}{4}$ & \hfil $\frac{1-W_2}{4}$ \\
\hfil $|A_1\ra\la A_2|, |A_2\ra\la A_3|$ & \hfil $\frac{1-W_1}{2}$ & \hfil 0 \\
\hfil $|A_0\ra\la A_2|, |A_2\ra\la A_4|$ & \hfil $\frac{1+W_2}{2}$ & \hfil 0 \\
\hfil $|A_1\ra\la A_3|$ & \hfil $\frac{1-2W_1+W_2}{4}$ & \hfil $\frac{1+2W_1+W_2}{4}$ \\
\hfil $|A_0\ra\la A_3|, |A_1\ra\la A_4|$ & \hfil $\frac{1-W_1+W_2-W_3}{4}$ & \hfil $\frac{1+W_1-W_2-W_3}{4}$ \\
\hfil $|A_0\ra\la A_4|$ & \hfil $\frac{1+W_2+W_4}{4}$ & \hfil $\frac{1-2W_2+W_4}{4}$ \\
\hline \hline
\end{tabular}
\label{tab:mat}
\end{table*}

When the ancillary photon pulse is successfully measured, i.e., the photon is not lost anywhere, we derive the density matrices according to the measurement results. The matrix elements of $|A_k\ra\la A_l|$ omitting the coefficients $\alpha_s$ are summarized in Table~\ref{tab:mat}.
%
\subsection{Twirling for numerical simulation}
\label{ssec:twir}
As shown in the last subsection, the noise maps for $N\geq 2$ cannot be described only by the Pauli channels, which indicates that they are not simulatable within the standard stabilizer-circuit simulation~\cite{aaronson2004} because of the exponential growth of their simulation cost with the number of qubits.
Therefore, in Sec.~\ref{sec:nume}, we calculate logical error rate by translating such non-trivial error maps to tractable ones through {\it Pauli twirling}~\cite{emerson2007,bendersky2008}.
In general, Pauli twirling somewhat changes the character of an original noise, thereby being debatable to be used for the direct comparison between experimental results and numerical simulation. However, in the case of stabilizer measurement, Pauli twirling can be performed even experimentally as shown below, which allows the direct comparison between them.

\begin{figure*}[t]
    \centering
    \includegraphics[width=16cm]{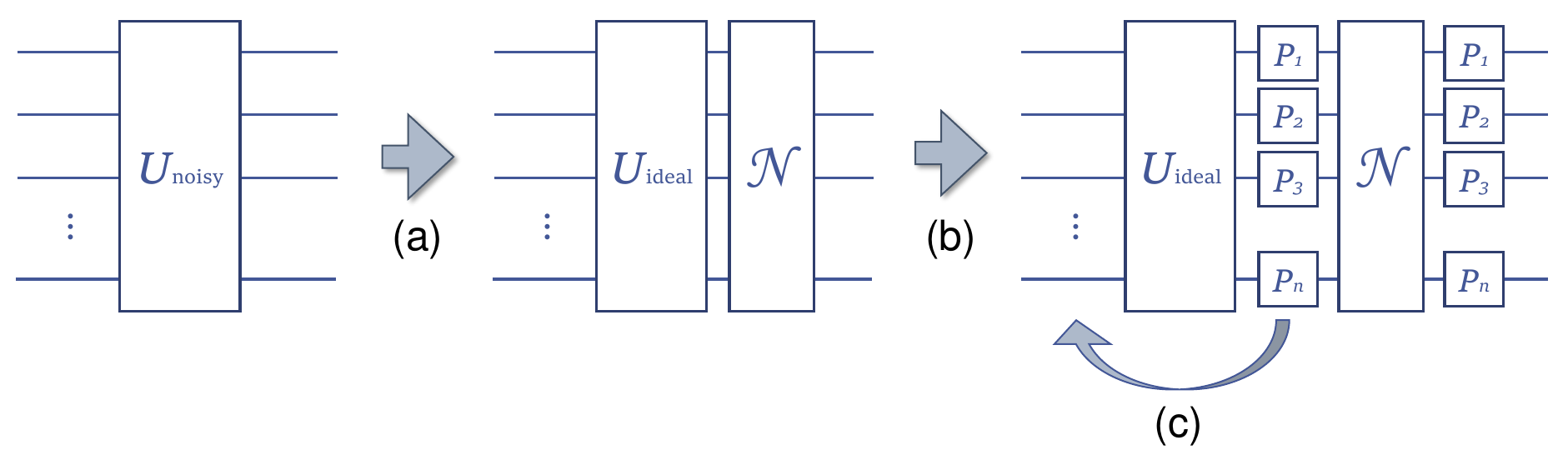}
    \caption{Schematic of Pauli twirling of a noisy operation.}
    \label{fig:twirling}
\end{figure*}

In the Pauli twirling, a noisy operation $U_{\rm noisy}$ is {\it virtually} decomposed into an ideal operation $U_{\rm ideal}$ and a remaining noise channel ((a) in Fig.~\ref{fig:twirling}). Then, a string of Pauli operations on every qubit is inserted immediately before and after the noise channel ((b) in Fig.~\ref{fig:twirling}). This converts the noise to Pauli errors. This is only a virtual process, but if the Pauli string immediately before the noise commutes the ideal operation, the insertion of Pauli strings immediately before and after the noisy operation is equivalent to the Pauli twirling of the noise. If the error rates of Pauli operations are assumed to be negligible compared to the two-qubit gates, this is feasible in experiments, which means that benchmarks in experiments can be directly compared with those in stabilizer simulation.
%
\section{Numerical analysis}
\label{sec:nume}
%
\subsection{Methods}
\label{ssec:method}
\subsubsection{procedure for calculating logical error}
\label{ssec:pro}
\begin{figure}[t]
    \centering
    \includegraphics[width=8.5cm]{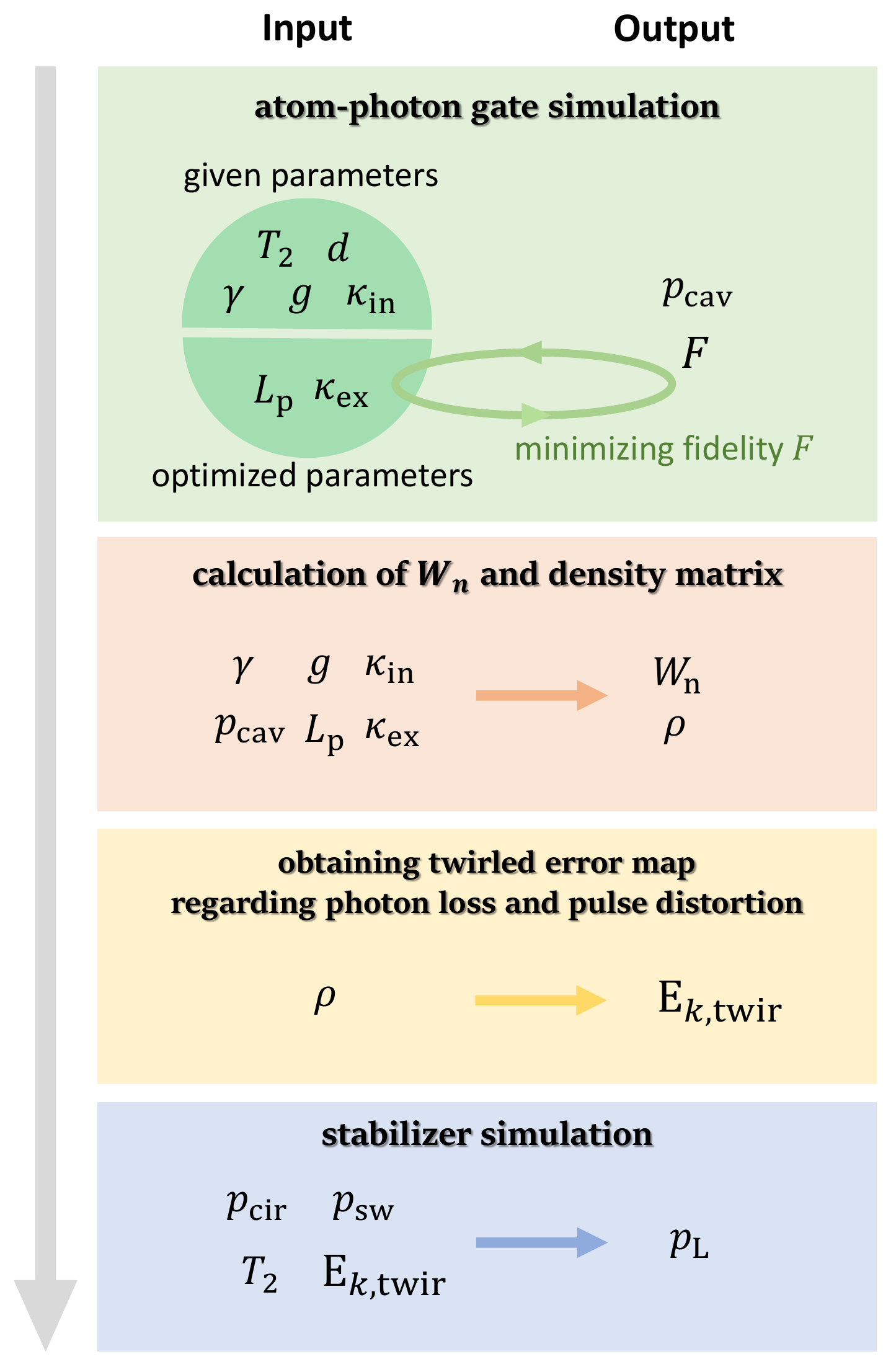}
    \caption{Summary of the procedure of the numerical analysis.}
    \label{fig:proc}
\end{figure}

This section explains a method to evaluate logical error rates according to CQED parameters.
In Fig.~\ref{fig:proc}, we provide a summary of the procedure of the numerical analysis. Input parameters and outputs are shown in each procedure.
First, we determine CQED configuration parameters to minimize the total infidelity caused by the errors during a syndrome measurement cycle as shown in Sec.~\ref{ssec:cpara}. From these parameters, we evaluate an effective noise map.
To simplify the simulation, we employ the following approximation.
\begin{enumerate}
\item We assume dephasing errors happen at the beginning of each syndrome-measurement cycle. 
\item the noise map of pulse-delay error is Pauli-twirled to a stochastic Pauli noise for efficient simulation. 
\item When photon loss happens in a certain cycle, we ignore the pulse-delay error of the cycle. 
\end{enumerate}

Next, we describe a detailed procedure for calculating the logical error rate.
In our simulation, each syndrome measurement outputs one among $0,1,2$, where $2$ means the corresponding photon was lost. Supposing we obtain $M$ syndrome values in total, we denote the sequence of syndrome values as $s \in \{0,1,2\}^M$ and define $\mathcal{E}_s$ as a completely positive map of noisy $d$ cycles when we obtain $s$ as the results of syndrome measurements. Then, the probability with which we obtain $s$ from input state $\rho$ is given by ${\rm Tr}[\mathcal{E}_s(\rho)]$. For technical reasons explained later, we assume the last cycle is error-free.
After the syndrome measurement, we estimate the Pauli recovery map $\mathcal{R}_s$ from a syndrome value $s$. We implemented two types of error estimation algorithms, which are explained in the next section. Thus, the final state after error correction is given by $\sum_s (\mathcal{R}_s \circ \mathcal{E}_s) (\rho)$. 

The logical error rates are calculated as the process fidelity of this channel. 
Since we assume that the first and last cycles are error-free, the whole channel is a logical channel, i.e., it maps any logical state to a logical state. Thus, the process fidelity can be evaluated as follows.
\begin{align}
f_{\rm pro} = \bra{\psi_{\rm Bell}} \sum_s (\mathcal{R}_s \circ \mathcal{E}_s) (\ket{\psi_{\rm Bell}}\bra{\psi_{\rm Bell}}) \ket{\psi_{\rm Bell}}
\end{align}
Here, $\ket{\psi_{\rm Bell}} = \frac{1}{2}(\ket{0}_{\rm L} \ket{0} + \ket{1}_{\rm L} \ket{1})$ is a Bell state of surface-code qubit and virtual qubit. The infidelity per cycle is calculated as $f_{\rm cyc} = (f_{\rm pro})^{1/d}$, and the logical error rate per cycle is given by $1-f_{\rm cyc}$.

As shown in Sec.~\ref{sec:erana}, all the noises considered in this paper can be treated as the probabilistic Pauli measurements and operations, and the initial Bell state is a stabilizer state. Thus, the logical error rates according to the CQED parameters can be simulated efficiently with the stabilizer formalism~\cite{aaronson2004}. To treat photon-loss events and correlated backactions due to pulse-delay errors, we create our own stabilizer simulators implemented with the C++ language. 
%
\subsubsection{Calculation of photon loss probability and $W_m$}
\label{ssec:cpara}
To calculate the photon loss probability for given sets of cavity parameters, we first simulate the dynamics of the CQED system. The Jaynes-Cummings Hamiltonian of our system is given by
\begin{align}
\nn
H &= \hbar\omc(\cldag c_l + \crdag c_r)
+ \hbar\ome|{\rm e}\ra\la{\rm e}| - \hbar\omg|{0}\ra\la{0}| \\
\nn
&+ \hbar\int_{-\infty}^{\infty}d\om\om(\ldag(\om)l(\om)+\rdag(\om)r(\om))  \\
\nn
&+ i\hbar g\lp(\crdag|1\ra\la{\rm e}| - c_r|{\rm e}\ra\la 1|\rp) \\
&+ i\hbar\sqrt{\frac{\kex}{\pi}}\int_{-\infty}^{\infty}d\om\lp[\cldag l(\om) - \ldag(\om)c_l + \crdag r(\om) - \rdag(\om)c_r\rp]   ,
\label{eq:ham}
\end{align}
where $c_s$ and $s(\om)$ ($s=l,r$) are the annihilation operators of a photon in a cavity and of a pulse photon for $L$- and $R$-polarization.
We also take into account the following dissipative processes: atomic spontaneous emission with a (polarization) decay rate $\gamma$, cavity field decay with a decay rate $\kex$ associated with the extraction of a cavity photon to the desired external mode via transmission of the mirror, and other undesirable cavity field decay due to the imperfection of the cavity with the rate $\kin$. The total cavity decay rate is given by ${\kappa=\kex+\kin}$.

Based on this model and the initial state given by Eq.~(\ref{eq:pin}), we solve the Lindblad master equation for the CQED system and calculate the photon loss probability and $W_m$ for the stabilizer simulation. Here we assume an input pulse has Gaussian shape.
In this calculation, the decay rate $\kex$ and pulse length $\len$ are adjustable parameters in an experiment, which has optimal values depending on applications.
In our case, $\kex$ and $\len$ are optimized to maximize the total fidelity, reflecting the deterioration due to all the error sources that we focus on in this paper.
Using given and optimized parameters, we calculate $W_m$ defined in Eq.~\ref{eq:wm} and noise map. When the pulse length is longer than the timescales of CQED dynamics, we can approximately calculate $W_m$ from
\begin{align}
\nonumber
W_m &= \frac{1}{\sqrt{\sqrt{\pi}\len}}\int^{\infty}_{-\infty} dt \exp \lp[-\frac{1}{2\len^2}(t^2+(t-m|\tau_0-\tau_1|))\rp] \\
&= \exp\lp[-\frac{m^2(\tau_0-\tau_1)^2}{4\len^2}\rp],
\label{eq:appwm}
\end{align}
where $\tau_0$ and $\tau_1$ are the pulse delay due to reflection by a cavity when an atom is in $|0\ra$ and $|1\ra$, respectively. They are given by
%
\begin{align}
\tau_0 &= \frac{2\kex}{\kex^2-\kin^2}, \\
\tau_1 &= \frac{2\kex(g^2-\gamma^2)}{g^4+\gamma^2(\kin^2-\kex^2)+2g^2\gamma\kin}.
\label{eq:appwm}
\end{align}
The details for the derivation of these equations are shown in Appendix~\ref{app:delay}.
%
\subsubsection{Error estimation algorithms}
\label{ssec:eea}
In this paper, logical error rates are estimated as follows. If there is no photon-loss event, i.e., $s \in \{0,1\}^M$, the error estimation of surface codes can be performed by converting the problem to a minimum-weight perfect matching~(MWPM) problem~\cite{fowler2012}, which can be efficiently solved using Edmonds' blossom algorithms~\cite{edmonds1965paths}. This conversion is based on the fact that any single error event will flip two syndrome values in surface codes. Thus, estimating one of the highest error patterns consistent with observed syndrome values is equal to choosing the minimum set of edges that will flip the observed syndrome values. If error rates are not uniform, we can reflect this in the error estimation by appropriately setting a non-uniform weight of edges, e.g., setting a small cost for high probability events.

Our error-decoding strategy in designing error estimation algorithms is to convert the error estimation tasks under photon losses to the above MWPM-based methods. We evaluate the performance under photon losses with two strategies: uniform decoding and weighted decoding.
In the uniform decoding, the syndrome value $2$ (loss) is replaced with the value observed in the last cycles. This is because the observed syndrome values are expected to be the same as the last cycle if error rates are small. Note that if it is the first cycle, $2$ is replaced with $0$. 
In the weighted decoding, the information of photon-loss events is adaptively utilized to reduce logical error rates. The loss of a syndrome-measurement photon erases syndrome values and adds Pauli errors to data qubits. Suppose a photon is expected to measure Pauli-$Z$ stabilizer on four data qubits named $(1,2,3,4)$, but the photon is lost after interacting with $i$-th data qubit. This is equal to performing Pauli $Z_1 \cdots Z_i$ measurement and forgetting the measurement results, i.e., randomly applying Pauli $Z_1 \cdots Z_i$ with probability $0.5$. 
Thus, the photon-loss signal indicates that the probabilities of Pauli errors on several subsets of target data qubits are much higher than the physical error probability. We can reflect this by adaptively setting the cost of corresponding edges small. There are three patterns of Pauli errors incurred by a photon loss, $Z_1$, $Z_1 Z_2$, and $Z_1 Z_2 Z_3$. We chose corresponding edges to these events and reduced the weight $\alpha$-times smaller according to the loss events, where $\alpha$ is a heuristically chosen real value. The same method is applied to the Pauli-$X$ stabilizer measurements.
While the weighted decoding performs better than the uniform decoding, its implementation is more challenging. In the execution of MWPM decoders, we need to calculate the shortest paths between two target nodes. Since we cannot predict when photon-loss events will happen, this path search in the weighted decoding must be performed at runtime, degrading the throughput of error estimation algorithms. If this becomes slower than the syndrome measurement cycles, we need to reduce the period of syndrome measurements from the optimal value, which increases effective physical error rates due to the $T_2$ lifetime. 
More aggressive utilization of photon-loss information might be possible, which we left as future work.
%
\subsection{Performance of CQED network required for FTQC}
\label{ssec:req}
\subsubsection{Fault-tolerance}
\label{ssec:tole}
\begin{figure*}[t]
    \centering
    \includegraphics[width=17.5cm]{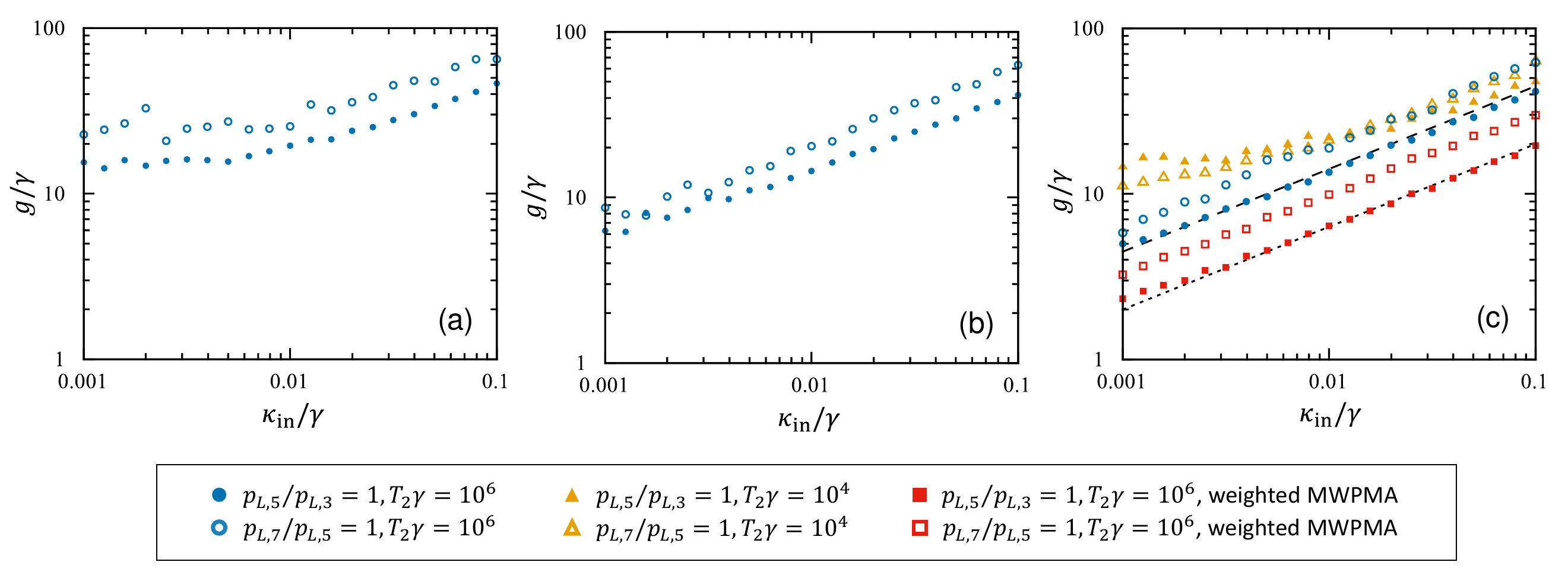}
    \caption{Requirements for a CQED system to reduce the logical error as the code size increases in (a) 4-cavity structure, (b) $d$-cavity structure, and (c) $N$-cavity structure. ${p_{\rm L,7}/p_{\rm L,5} < 1}$ and ${p_{\rm L,5}/p_{\rm L,3} < 1}$ for lager values of $g/\gamma$ than open and filled symbols, respectively. In (c), squares indicate the requirement in the case of the improved error estimation making effective use of the photon-loss information at the photodetectors. Dotted and dashed lines denote ${\cin=2000}$ and 10000, respectively. The sample size in the Monte Carlo simulation for calculating logical error rates is $10^5$ (the same applies to all the simulations in the subsequent figures). Here, we assume that the peripheral devices are ideal, namely, $\psw+\pci=0$.}
    \label{fig:thre_cp}
\end{figure*}

\begin{figure}[t]
    \centering
    \includegraphics[width=8.5cm]{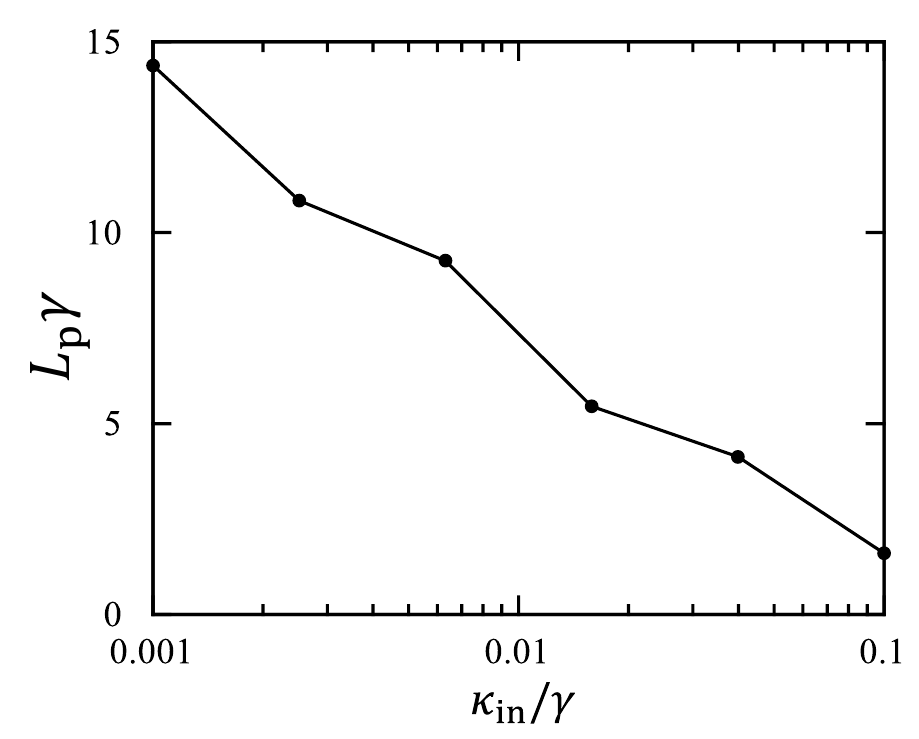}
    \caption{Pulse length for $d=5$ along the requirement boundary ${p_{\rm L,5}/p_{\rm L,3} = 1}$ in the 4-cavity structure.}
    \label{fig:pullen}
\end{figure}

 Figure~\ref{fig:thre_cp}(a)-(c) show requirements for a CQED system to reduce the logical error as the code size increases for the 4-cavity structure, $d$-cavity structure, and $N$-cavity structure, respectively. In the region above the requirement boundaries, $p_{\rm L,5(7)}$ is lower than $p_{\rm L,3(5)}$. We calculated these results using the MWPM algorithm with $10^5$ samples in Monte Carlo simulation.
For the same value of $T_2$, the requirement for a CQED system is more strict in a structure with fewer cavities because it has low parallelism, resulting in strong suffering from the dephasing error. This is pronounced in the region where $\kin/\gamma$ is small. For small $\kin$, the line width of a cavity is narrow, which prevents short input pulses. This indicates that an input pulse for small $\kin$ must be long, and the low parallelism strongly deteriorates atomic qubits due to the dephasing error. In Fig.~\ref{fig:pullen}, we plot the pulse length along the requirement boundary ${p_{\rm L,5}/p_{\rm L,3} = 1}$ in the 4-cavity structure when the code distance is 5. The pulse length is longer when the value of $\kin$ is smaller.
A short $T_2$ also makes the requirement strict, particularly for small $\kin$. Figure~\ref{fig:thre_cp}(c) shows the difference between the requirements for $T_2\gamma=10^4$ and $T_2\gamma=10^6$. This is also caused by the narrow line width of a cavity preventing short input pulses.

Here, as mentioned in Sec.~\ref{ssec:est}, the requirement has a threshold value for FTQC only in the case of the $N$-cavity structure, at least within our noise model. That is, the 4- and $d$-cavity structures do not have a finite threshold value in the sense that the error rate per gate operation increases as the code distance increases. This means that these structures experience the increase in the logical error rate as the code distance increases over a certain code distance.
Nevertheless, considering such structures is helpful to implement high-performance quantum computer. In realistic architectures using any physical platforms, achieving a system that has a finite threshold value in the strict sense is likely to be a somewhat daunting challenge; thus, finding a compromise architecture that does not have the threshold but can still have sufficient fault-tolerance in practice will be an avoidable issue for the future, and therefore we also explore such architectures in this paper.

\begin{figure}[t]
    \centering
    \includegraphics[width=8.5cm]{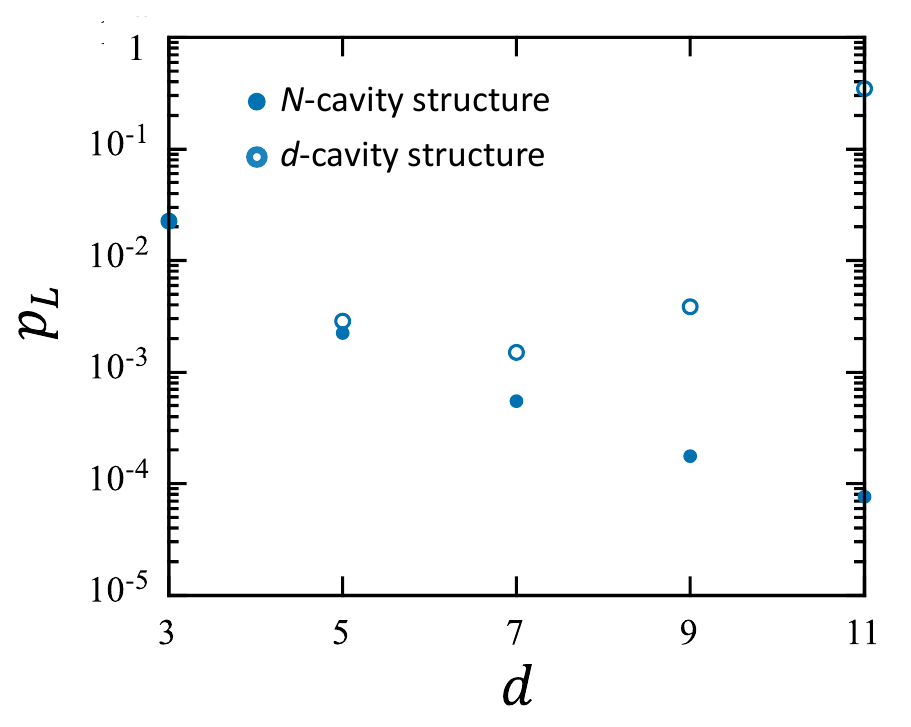}
    \caption{Logical error rates for the $d$-cavity structure (open circles) and the $N$-cavity structure (filled circles) as a function of the code distance $d$. The cavity parameters are $(g/\gamma,\kin/\gamma)=(50.12,0.01)$ for both structures.}
    \label{fig:ddep}
\end{figure}

In Fig.~\ref{fig:ddep}, we plot the code-distance dependence of the logical error rate in $N$- and $d$-cavity structures for a cavity parameter set that satisfies the fault-tolerance of the $N$-cavity structure.
The figure indicates that while the logical error rate in the $N$-cavity structure monotonously decreases as the code distance $d$, the one in the $d$-cavity structure rebounds when the code distance exceeds 7.
This means that choosing a CQED-network structure involves a trade-off between fault-tolerance and implementation difficulty. Thus, in building a quantum computer, one should choose a reasonable structure that can achieve a logical error rate required for excusing a quantum algorithm with high precision, suppressing implementation difficulty.
Figure~\ref{fig:ddep} does not include the result of the 4-cavity structure since it is difficult to plot the data alongside those of the other structures in easy-to-see format for the same CQED parameter set, but the 4-cavity structure also shows the rebound behavior as well as the $d$-cavity structure.
%
\subsubsection{Improvement of error estimation using photon-loss information}
\label{ssec:imp}
In our results, the requirement for the CQED parameters is severe, e.g., the boundary of ${p_{\rm L,5}/p_{\rm L,3} = 1}$ requires $\cin = 10000$, where $\cin\equiv \frac{g^2}{2\kin\gamma}$, a fundamental characteristic of the cavity performance for quantum computing~\cite{goto2019,asaoka2021,schupp2021,utsugi2025}, whereas the value of the internal cooperativity in experiments is in the order of 100 at best~\cite{gehr2010,hunger2010,ruddell2020}.
On the other hand, we find that the weighted decoding, an improved error estimation method utilizing the photon-loss information at the photodetectors introduced in Sec.~\ref{ssec:eea}, relaxes this requirement significantly. In Fig.~\ref{fig:thre_cp}(c), we can see that the value of the internal cooperativity achieving ${p_{\rm L,5}/p_{\rm L,3} = 1}$ is 1/5 compared to the case of using the normal MWPM algorithm. This requirement is still high for realizing the FTQC with a CQED network. Nevertheless, considering room for improving the MWPM algorithm or error-correcting code itself to be more suitable for the CQED-network-based quantum computing, our results give the motivation to boost studies in this field.
%
\subsubsection{Influence of loss at peripheral devices}
\label{ssec:otherloss}
\begin{figure}[t]
    \centering
    \includegraphics[width=8.5cm]{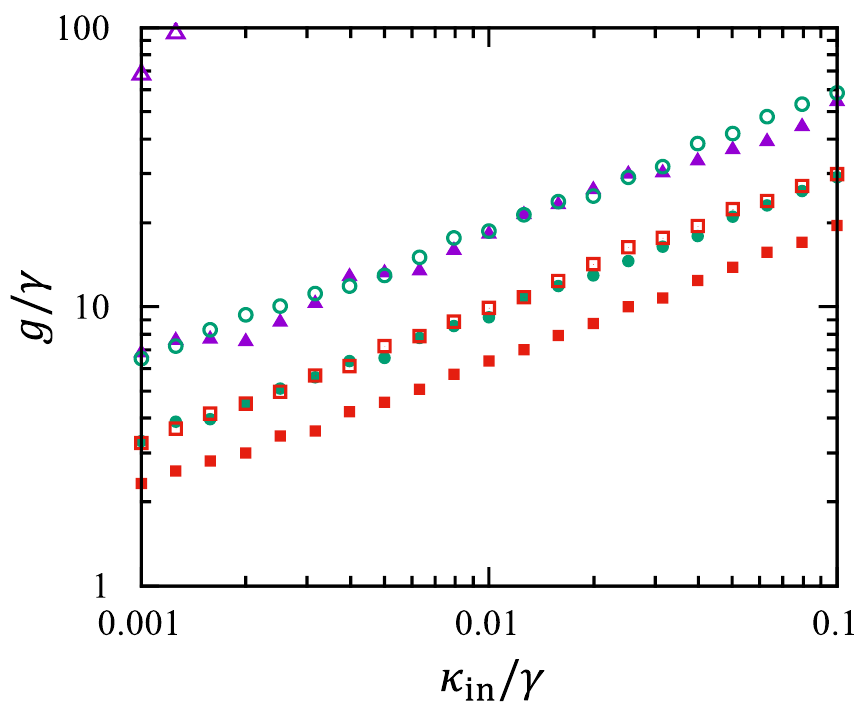}
    \caption{Requirements for CQED parameters to reduce the logical error as the code size increases when the total loss rate of the peripheral devices is ${\psw+\pci=0.01}$(circles), 0.02(triangles), and 0(squares), namely, the same as the result in Fig.~\ref{fig:thre_cp}(c). Open and filled symbols represent the points at which ${p_{\rm L,7}/p_{\rm L,5} = 1}$ and ${p_{\rm L,5}/p_{\rm L,3} = 1}$, respectively. Simulations here are performed based on the weighted decoding. Dephasing time is set to ${T_2\gamma = 10^6}$.}
    \label{fig:lossdep_thre}
\end{figure}

Here, we comment on the influence of photon loss on peripheral devices, i.e., optical switches and circulators which route ancillary photons. Figure~\ref{fig:lossdep_thre} shows the dependence of the requirement boundary on the total photon loss rate at switches and circulators ${\psw+\pci}$, simulated based on the weighted decoding. The requirement becomes drastically demanding when the total photon loss rate exceeds $~2\%$. This indicates that the entire photon loss rate in the CQED networks is at most $~2\%$ to achieve FTQC. Although this loss requirement is quite high for commercially available optical switches and circulators, recent development of photon-routing technologies using cavity- or waveguide-QED has the potential to achieve this requirement~\cite{aoki2009,shomroni2014,scheucher2016,papon2019,ren2022}. Moreover, our structure is still straightforward and has room for improving robustness to photon loss.
\begin{figure}[t]
    \centering
    \includegraphics[width=8.5cm]{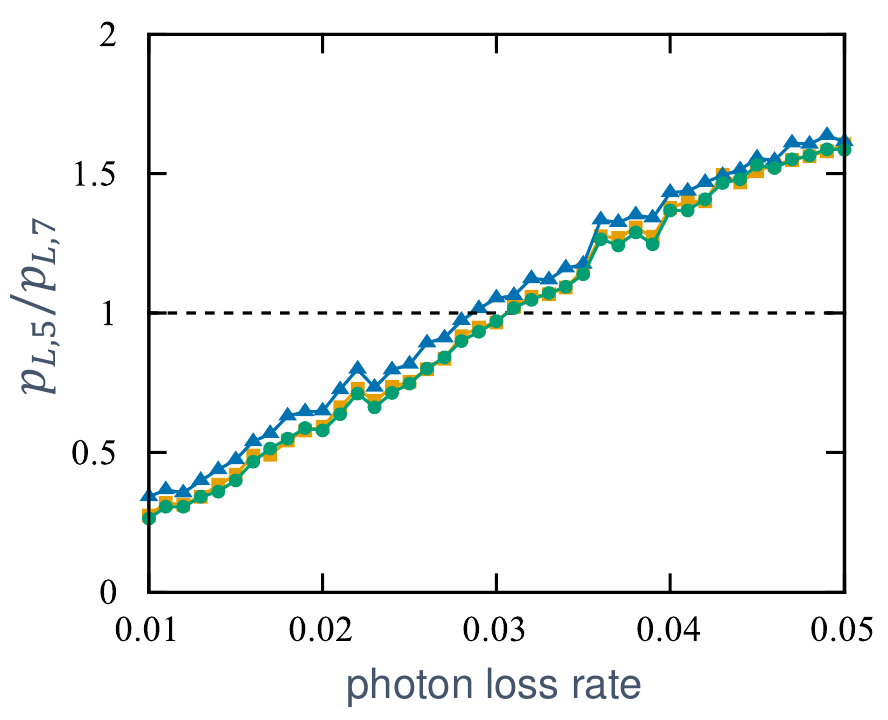}
    \caption{Ratio of $p_{L,5}$ to $p_{L,7}$ as a function of the total photon loss rate, including the cavity loss and the losses at the peripheral devices. We set the infidelity of the CZ gate $1-F=10^{-3}$(circles), $5\times 10^{-4}$(squares), and $10^{-4}$(triangles) and sweep the total photon loss rate regardless of the cavity parameters. The dashed line denotes ${p_{L,5}/p_{L,7}=1}$.}
    \label{fig:lossdep_logi}
\end{figure}
%
\section{Discussion and Conclusion}
\label{sec:discon}
In this paper, we have investigated the architecture of fault-tolerant quantum computing with neutral atoms, where atoms are trapped with CQED networks. We provided a concrete description of the architecture and proposed several network structures that have some advantages compared to the existing ones. Their performance was numerically evaluated and compared based on practical noise models. 

As this paper focused on providing a first stepping-stone architecture for developing CQED-network-based FTQC, there would be several directions to maximize the performance of CQED-network-based architecture: The properties of cavities and atoms could be tuned for optimizing the performance of logical error rates. Some contrivances may improve the parallelism of operations while also considering the possibility of combining them with atomic-array technologies. As one of the most natural ways, Ones of the most promising approaches are to explore more suitable quantum error-correcting codes that can take advantage of nearly all-to-all connectivity in CQED-network systems or that is tailored for realistic noise properties as done in the case of biased noise~\cite{aliferis2008,tuckett2018,tuckett2020,darmawan2021,wu2022,claes2023,sahay2023}. Furthermore, the performance of our architecture is sensitive to the performance of optical switches and circulators. Thus, our results also motivate the exploration of low-loss single-photon-level switches and circulators.

When we utilize a nano-fiber cavity structure, we can trap multiple atoms for each cavity and select the qubit diagnosed by an incident photon. Since internal atoms can interact, for example, with Rydberg interactions, we can utilize 2.5D architecture to perform logical CNOT with transversal CNOT gates as discussed in the idea of virtualized logical qubits~\cite{duckering2020,ramette2022}. Since the structure of fiber networks can be designed flexibly, further optimization may be possible.

Extending our architecture to distributed quantum computing is also a promising direction. While the number of cavities in a single vacuum chamber would be limited, fiber-coupled photons can be easily transported to another chamber. This is an obvious advantage compared to the architecture with microwave photons.
%
\section*{Acknowledgement}
This work was supported by JST Moonshot R\&D Grant Number JPMJMS2061 and JPMJMS2268, JST CREST Grant Number JPMJCR1771, JPMJCR23I4, and JPMJCR24I4, MEXT Q-LEAP Grant Number JPMXS0120319794.

\appendix
\section{Treatment of pulse-distortion error}
\label{app:delay}
This appendix explains the mathematical treatment of the pulse-distortion error in the atom-photon controlled-Z gate.
The Hamiltonian of the present CQED system is given by
\begin{align}
\nn
H &= \hbar\omc \cdag c
+ \hbar\ome|{\rm e}\ra\la{\rm e}| - \hbar\omg|{0}\ra\la{0}| \\
\nn
&+ \hbar\int_{-\infty}^{\infty}d\om\om\adag(\om)a(\om)  \\
\nn
&+ i\hbar g\lp(\cdag|1\ra\la{\rm e}| - c|{\rm e}\ra\la 1|\rp) \\
&+ i\hbar\sqrt{\frac{\kex}{\pi}}\int_{-\infty}^{\infty}d\om\lp[\cdag a(\om) - \adag(\om)c\rp]   ,
\label{eq:ham}
\end{align}
where $c (\cdag)$ and $a (\adag)$ are the annihilation (creation) operators of a photon in the cavity and of a pulse photon, respectively. We also take into account the following dissipative processes: atomic spontaneous emission with a (polarization) decay rate $\gamma$, cavity field decay with a decay rate $\kex$ associated with the extraction of a cavity photon to the desired external mode via transmission of the mirror, and other undesirable cavity field decay due to the imperfection of the cavity with the rate $\kin$. The total cavity decay rate is given by ${\kappa=\kex+\kin}$.

We consider that the single-photon pulse is nearly resonant with the cavity, and we assume that the atomic state is approximately independent of time. From the Hamiltonian~(\ref{eq:ham}), we derive response functions, namely the amplitude reflection coefficient of the photon pulse defined by $F_{q, {\rm out}} =L_q(\Delta)F_{q, {\rm in}}$ with the probability amplitude of the input (output) pulse at the position of the input mirror when the atom is in $|q\ra$, $F_{q, {\rm in(out)}}$ (see Ref.~\cite{asaoka2021,utsugi2025} for the detail):
\begin{align}
L_0(\Delta) &= \frac{-\kex + \kin - i\Delta}{\kex + \kin - i\Delta}   ,
\label{eq:l0}
\\
L_1(\Delta) &= \frac{-\kex + \kin -i\Delta + \frac{g^2}{\gamma-i\Delta}}{\kex + \kin -i\Delta + \frac{g^2}{\gamma-i\Delta}}   ,
\label{eq:l1}
\end{align}
where ${\Delta = \om - \omc}$ is the detuning. To realize the CPF gate, only the phase of the atomic state $|0\ra$ needs to change by $\pi$, namely ${L_0 = -1}$ and ${L_1 = 1}$. We expand Eqs.~(\ref{eq:l0}) and (\ref{eq:l1}) by $\Delta$ as
\begin{align}
\nn
L_0(\Delta) &= 1 - \frac{2\kex}{\kappa}
+ i\frac{2\kex}{\kappa^2}\Delta
+ \frac{2\kex}{\kappa^3}\Delta^2
+ O(\Delta^3),
\label{eq:l0ex}
\\
L_1(\Delta) &= 1 - \frac{2\kex\gamma}{g^2+\kappa\gamma}
- i\frac{2\kex(g^2-\gamma^2)}{(g^2+\kappa\gamma)^2}\Delta \\
&\hspace{1cm}- \frac{2\kex(2g^2\gamma+g^2\gamma-\gamma^3)}{(g^2+\kappa\gamma)^3}\Delta^2
+ O(\Delta^3).
\label{eq:l1ex}
\end{align}
Then, we neglect the second and higher-order terms. This approximation is roughly justified when the pulse duration $W$ satisfies $W\gg\max(1/\kappa,\kappa/g^2)$, which is derived from the condition that the first-order terms are sufficiently smaller than the second-order terms. Within the first-order approximation, the response functions represent norm reduction (i.e., photon loss) and pulse delay as follows.
Let the coefficients of the first order terms in the response functions (\ref{eq:l0ex}) and (\ref{eq:l1ex}) be $L'_0$ and $L'_1$, respectively, i.e., ${L_{0,1}(\Delta)=L_{0,1}(0)+L'_{0,1}\Delta}$ after performing the first order approximation. These functions are approximately rewritten as ${L_{0,1}(\Delta)\simeq L_{0,1}(0)e^{-i\tau_{0,1}\Delta}}$. Here we have defined the ${-i\tau_{0,1}\equiv L'_{0,1}/L_{0,1}(0)}$. From the time shifting property of Fourier transformation ${\bold F}$ represented by ${\bold F}[f(t-a)](k)={\bold F}[f(t)](\Delta)e^{-iak}$, the relation between the wavefunctions of the output and the input pulse $f_{0,1}^{\rm in,out}(t)$ is given by
\begin{align}
\nn
\fout(t)
&= {\bold F}^{-1}[{\bold F}[\fin(t)](\Delta)L_{0,1}(\Delta)]   \\
\nn
&= L_{0,1}(0){\bold F}^{-1}[{\bold F}[f_{0,1}(t)^{\rm in}](\Delta)e^{-i\tau_{0,1}\Delta}]   \\
&= L_{0,1}(0)\fin(t-\tau_{0,1}).
\label{eq:fout}
\end{align}
Thus, the first-order approximation reduces the error model to the norm reduction and the pulse delay.

\bibliographystyle{apsrev4-2}
\bibliography{cite_asaoka}

\begin{thebibliography}{70}%
\makeatletter
\providecommand \@ifxundefined [1]{%
 \@ifx{#1\undefined}
}%
\providecommand \@ifnum [1]{%
 \ifnum #1\expandafter \@firstoftwo
 \else \expandafter \@secondoftwo
 \fi
}%
\providecommand \@ifx [1]{%
 \ifx #1\expandafter \@firstoftwo
 \else \expandafter \@secondoftwo
 \fi
}%
\providecommand \natexlab [1]{#1}%
\providecommand \enquote  [1]{``#1''}%
\providecommand \bibnamefont  [1]{#1}%
\providecommand \bibfnamefont [1]{#1}%
\providecommand \citenamefont [1]{#1}%
\providecommand \href@noop [0]{\@secondoftwo}%
\providecommand \href [0]{\begingroup \@sanitize@url \@href}%
\providecommand \@href[1]{\@@startlink{#1}\@@href}%
\providecommand \@@href[1]{\endgroup#1\@@endlink}%
\providecommand \@sanitize@url [0]{\catcode `\\12\catcode `\$12\catcode
  `\&12\catcode `\#12\catcode `\^12\catcode `\_12\catcode `\%12\relax}%
\providecommand \@@startlink[1]{}%
\providecommand \@@endlink[0]{}%
\providecommand \url  [0]{\begingroup\@sanitize@url \@url }%
\providecommand \@url [1]{\endgroup\@href {#1}{\urlprefix }}%
\providecommand \urlprefix  [0]{URL }%
\providecommand \Eprint [0]{\href }%
\providecommand \doibase [0]{https://doi.org/}%
\providecommand \selectlanguage [0]{\@gobble}%
\providecommand \bibinfo  [0]{\@secondoftwo}%
\providecommand \bibfield  [0]{\@secondoftwo}%
\providecommand \translation [1]{[#1]}%
\providecommand \BibitemOpen [0]{}%
\providecommand \bibitemStop [0]{}%
\providecommand \bibitemNoStop [0]{.\EOS\space}%
\providecommand \EOS [0]{\spacefactor3000\relax}%
\providecommand \BibitemShut  [1]{\csname bibitem#1\endcsname}%
\let\auto@bib@innerbib\@empty
\bibitem [{\citenamefont {Shor}(1997)}]{shor1997}%
  \BibitemOpen
  \bibfield  {author} {\bibinfo {author} {\bibfnamefont {P.~W.}\ \bibnamefont
  {Shor}},\ }\href@noop {} {\bibfield  {journal} {\bibinfo  {journal} {SIAM J.
  Computing}\ }\textbf {\bibinfo {volume} {26}},\ \bibinfo {pages} {1484}
  (\bibinfo {year} {1997})}\BibitemShut {NoStop}%
\bibitem [{\citenamefont {Gidney}\ and\ \citenamefont
  {Eker{\aa}}(2021)}]{gidney2021}%
  \BibitemOpen
  \bibfield  {author} {\bibinfo {author} {\bibfnamefont {C.}~\bibnamefont
  {Gidney}}\ and\ \bibinfo {author} {\bibfnamefont {M.}~\bibnamefont
  {Eker{\aa}}},\ }\href@noop {} {\bibfield  {journal} {\bibinfo  {journal}
  {Quantum}\ }\textbf {\bibinfo {volume} {5}},\ \bibinfo {pages} {433}
  (\bibinfo {year} {2021})}\BibitemShut {NoStop}%
\bibitem [{\citenamefont {Harrow}\ \emph {et~al.}(2009)\citenamefont {Harrow},
  \citenamefont {Hassidim},\ and\ \citenamefont {Lloyd}}]{harrow2009}%
  \BibitemOpen
  \bibfield  {author} {\bibinfo {author} {\bibfnamefont {A.~W.}\ \bibnamefont
  {Harrow}}, \bibinfo {author} {\bibfnamefont {A.}~\bibnamefont {Hassidim}},\
  and\ \bibinfo {author} {\bibfnamefont {S.}~\bibnamefont {Lloyd}},\
  }\href@noop {} {\bibfield  {journal} {\bibinfo  {journal} {Phys. Rev. Lett.}\
  }\textbf {\bibinfo {volume} {103}},\ \bibinfo {pages} {150502} (\bibinfo
  {year} {2009})}\BibitemShut {NoStop}%
\bibitem [{\citenamefont {Babbush}\ \emph {et~al.}(2018)\citenamefont
  {Babbush}, \citenamefont {Gidney}, \citenamefont {Berry}, \citenamefont
  {Wiebe}, \citenamefont {McClean}, \citenamefont {Paler}, \citenamefont
  {Fowler},\ and\ \citenamefont {Neven}}]{babbush2018}%
  \BibitemOpen
  \bibfield  {author} {\bibinfo {author} {\bibfnamefont {R.}~\bibnamefont
  {Babbush}}, \bibinfo {author} {\bibfnamefont {C.}~\bibnamefont {Gidney}},
  \bibinfo {author} {\bibfnamefont {D.~W.}\ \bibnamefont {Berry}}, \bibinfo
  {author} {\bibfnamefont {N.}~\bibnamefont {Wiebe}}, \bibinfo {author}
  {\bibfnamefont {J.}~\bibnamefont {McClean}}, \bibinfo {author} {\bibfnamefont
  {A.}~\bibnamefont {Paler}}, \bibinfo {author} {\bibfnamefont
  {A.}~\bibnamefont {Fowler}},\ and\ \bibinfo {author} {\bibfnamefont
  {H.}~\bibnamefont {Neven}},\ }\href@noop {} {\bibfield  {journal} {\bibinfo
  {journal} {Phys. Rev. X}\ }\textbf {\bibinfo {volume} {8}},\ \bibinfo {pages}
  {041015} (\bibinfo {year} {2018})}\BibitemShut {NoStop}%
\bibitem [{\citenamefont {Kivlichan}\ \emph {et~al.}(2020)\citenamefont
  {Kivlichan}, \citenamefont {Gidney}, \citenamefont {Berry}, \citenamefont
  {Wiebe}, \citenamefont {McClean}, \citenamefont {Sun}, \citenamefont {Jiang},
  \citenamefont {Rubin}, \citenamefont {Fowler}, \citenamefont {Aspuru-Guzik}
  \emph {et~al.}}]{kivlichan2020}%
  \BibitemOpen
  \bibfield  {author} {\bibinfo {author} {\bibfnamefont {I.~D.}\ \bibnamefont
  {Kivlichan}}, \bibinfo {author} {\bibfnamefont {C.}~\bibnamefont {Gidney}},
  \bibinfo {author} {\bibfnamefont {D.~W.}\ \bibnamefont {Berry}}, \bibinfo
  {author} {\bibfnamefont {N.}~\bibnamefont {Wiebe}}, \bibinfo {author}
  {\bibfnamefont {J.}~\bibnamefont {McClean}}, \bibinfo {author} {\bibfnamefont
  {W.}~\bibnamefont {Sun}}, \bibinfo {author} {\bibfnamefont {Z.}~\bibnamefont
  {Jiang}}, \bibinfo {author} {\bibfnamefont {N.}~\bibnamefont {Rubin}},
  \bibinfo {author} {\bibfnamefont {A.}~\bibnamefont {Fowler}}, \bibinfo
  {author} {\bibfnamefont {A.}~\bibnamefont {Aspuru-Guzik}}, \emph {et~al.},\
  }\href@noop {} {\bibfield  {journal} {\bibinfo  {journal} {Quantum}\ }\textbf
  {\bibinfo {volume} {4}},\ \bibinfo {pages} {296} (\bibinfo {year}
  {2020})}\BibitemShut {NoStop}%
\bibitem [{\citenamefont {Shor}(1995)}]{shor1995}%
  \BibitemOpen
  \bibfield  {author} {\bibinfo {author} {\bibfnamefont {P.~W.}\ \bibnamefont
  {Shor}},\ }\href@noop {} {\bibfield  {journal} {\bibinfo  {journal} {Phys.
  Rev. A}\ }\textbf {\bibinfo {volume} {52}},\ \bibinfo {pages} {R2493}
  (\bibinfo {year} {1995})}\BibitemShut {NoStop}%
\bibitem [{\citenamefont {Shor}(1996)}]{shor1996}%
  \BibitemOpen
  \bibfield  {author} {\bibinfo {author} {\bibfnamefont {P.~W.}\ \bibnamefont
  {Shor}},\ }\href@noop {} {\bibfield  {journal} {\bibinfo  {journal}
  {Proceedings of 37th Conference on Foundations of Computer Science}\ ,\
  \bibinfo {pages} {56}} (\bibinfo {year} {1996})}\BibitemShut {NoStop}%
\bibitem [{\citenamefont {Calderbank}\ and\ \citenamefont
  {Shor}(1996)}]{calderbank1996}%
  \BibitemOpen
  \bibfield  {author} {\bibinfo {author} {\bibfnamefont {A.~R.}\ \bibnamefont
  {Calderbank}}\ and\ \bibinfo {author} {\bibfnamefont {P.~W.}\ \bibnamefont
  {Shor}},\ }\href@noop {} {\bibfield  {journal} {\bibinfo  {journal} {Phys.
  Rev. A}\ }\textbf {\bibinfo {volume} {54}},\ \bibinfo {pages} {1098}
  (\bibinfo {year} {1996})}\BibitemShut {NoStop}%
\bibitem [{\citenamefont {Saffman}\ \emph {et~al.}(2010)\citenamefont
  {Saffman}, \citenamefont {Walker},\ and\ \citenamefont
  {M\o{}lmer}}]{saffman2010}%
  \BibitemOpen
  \bibfield  {author} {\bibinfo {author} {\bibfnamefont {M.}~\bibnamefont
  {Saffman}}, \bibinfo {author} {\bibfnamefont {T.~G.}\ \bibnamefont
  {Walker}},\ and\ \bibinfo {author} {\bibfnamefont {K.}~\bibnamefont
  {M\o{}lmer}},\ }\href@noop {} {\bibfield  {journal} {\bibinfo  {journal}
  {Rev. Mod. Phys.}\ }\textbf {\bibinfo {volume} {82}},\ \bibinfo {pages}
  {2313} (\bibinfo {year} {2010})}\BibitemShut {NoStop}%
\bibitem [{\citenamefont {Saffman}(2016)}]{saffman2016}%
  \BibitemOpen
  \bibfield  {author} {\bibinfo {author} {\bibfnamefont {M.}~\bibnamefont
  {Saffman}},\ }\href@noop {} {\bibfield  {journal} {\bibinfo  {journal} {J.
  Phys. B}\ }\textbf {\bibinfo {volume} {49}},\ \bibinfo {pages} {202001}
  (\bibinfo {year} {2016})}\BibitemShut {NoStop}%
\bibitem [{\citenamefont {Wang}\ \emph {et~al.}(2016)\citenamefont {Wang},
  \citenamefont {Kumar}, \citenamefont {Wu},\ and\ \citenamefont
  {Weiss}}]{wang2016}%
  \BibitemOpen
  \bibfield  {author} {\bibinfo {author} {\bibfnamefont {Y.}~\bibnamefont
  {Wang}}, \bibinfo {author} {\bibfnamefont {A.}~\bibnamefont {Kumar}},
  \bibinfo {author} {\bibfnamefont {T.-Y.}\ \bibnamefont {Wu}},\ and\ \bibinfo
  {author} {\bibfnamefont {D.~S.}\ \bibnamefont {Weiss}},\ }\href@noop {}
  {\bibfield  {journal} {\bibinfo  {journal} {Science}\ }\textbf {\bibinfo
  {volume} {352}},\ \bibinfo {pages} {1562} (\bibinfo {year}
  {2016})}\BibitemShut {NoStop}%
\bibitem [{\citenamefont {Levine}\ \emph {et~al.}(2018)\citenamefont {Levine},
  \citenamefont {Keesling}, \citenamefont {Omran}, \citenamefont {Bernien},
  \citenamefont {Schwartz}, \citenamefont {Zibrov}, \citenamefont {Endres},
  \citenamefont {Greiner} \emph {et~al.}}]{levine2018}%
  \BibitemOpen
  \bibfield  {author} {\bibinfo {author} {\bibfnamefont {H.}~\bibnamefont
  {Levine}}, \bibinfo {author} {\bibfnamefont {A.}~\bibnamefont {Keesling}},
  \bibinfo {author} {\bibfnamefont {A.}~\bibnamefont {Omran}}, \bibinfo
  {author} {\bibfnamefont {H.}~\bibnamefont {Bernien}}, \bibinfo {author}
  {\bibfnamefont {S.}~\bibnamefont {Schwartz}}, \bibinfo {author}
  {\bibfnamefont {A.~S.}\ \bibnamefont {Zibrov}}, \bibinfo {author}
  {\bibfnamefont {M.}~\bibnamefont {Endres}}, \bibinfo {author} {\bibfnamefont
  {M.}~\bibnamefont {Greiner}}, \emph {et~al.},\ }\href@noop {} {\bibfield
  {journal} {\bibinfo  {journal} {Phys. Rev. Lett.}\ }\textbf {\bibinfo
  {volume} {121}},\ \bibinfo {pages} {123603} (\bibinfo {year}
  {2018})}\BibitemShut {NoStop}%
\bibitem [{\citenamefont {Madjarov}\ \emph {et~al.}(2020)\citenamefont
  {Madjarov}, \citenamefont {Covey}, \citenamefont {Shaw}, \citenamefont
  {Choi}, \citenamefont {Kale}, \citenamefont {Cooper}, \citenamefont
  {Pichler}, \citenamefont {Schkolnik} \emph {et~al.}}]{madjarov2020}%
  \BibitemOpen
  \bibfield  {author} {\bibinfo {author} {\bibfnamefont {I.~S.}\ \bibnamefont
  {Madjarov}}, \bibinfo {author} {\bibfnamefont {J.~P.}\ \bibnamefont {Covey}},
  \bibinfo {author} {\bibfnamefont {A.~L.}\ \bibnamefont {Shaw}}, \bibinfo
  {author} {\bibfnamefont {J.}~\bibnamefont {Choi}}, \bibinfo {author}
  {\bibfnamefont {A.}~\bibnamefont {Kale}}, \bibinfo {author} {\bibfnamefont
  {A.}~\bibnamefont {Cooper}}, \bibinfo {author} {\bibfnamefont
  {H.}~\bibnamefont {Pichler}}, \bibinfo {author} {\bibfnamefont
  {V.}~\bibnamefont {Schkolnik}}, \emph {et~al.},\ }\href@noop {} {\bibfield
  {journal} {\bibinfo  {journal} {Nat. Phys.}\ }\textbf {\bibinfo {volume}
  {16}},\ \bibinfo {pages} {857^^e2^^80^^93861} (\bibinfo {year}
  {2020})}\BibitemShut {NoStop}%
\bibitem [{\citenamefont {Browaeys}\ and\ \citenamefont
  {Lahaye}(2020)}]{browaeys2020}%
  \BibitemOpen
  \bibfield  {author} {\bibinfo {author} {\bibfnamefont {A.}~\bibnamefont
  {Browaeys}}\ and\ \bibinfo {author} {\bibfnamefont {T.}~\bibnamefont
  {Lahaye}},\ }\href@noop {} {\bibfield  {journal} {\bibinfo  {journal} {Nat.
  Phys.}\ }\textbf {\bibinfo {volume} {16}},\ \bibinfo {pages} {132} (\bibinfo
  {year} {2020})}\BibitemShut {NoStop}%
\bibitem [{\citenamefont {Graham}\ \emph {et~al.}(2022)\citenamefont {Graham},
  \citenamefont {Song}, \citenamefont {Scott}, \citenamefont {Poole},
  \citenamefont {Phuttitarn}, \citenamefont {Jooya}, \citenamefont {Eichler},
  \citenamefont {Jiang} \emph {et~al.}}]{graham2022}%
  \BibitemOpen
  \bibfield  {author} {\bibinfo {author} {\bibfnamefont {T.~M.}\ \bibnamefont
  {Graham}}, \bibinfo {author} {\bibfnamefont {Y.}~\bibnamefont {Song}},
  \bibinfo {author} {\bibfnamefont {J.}~\bibnamefont {Scott}}, \bibinfo
  {author} {\bibfnamefont {C.}~\bibnamefont {Poole}}, \bibinfo {author}
  {\bibfnamefont {L.}~\bibnamefont {Phuttitarn}}, \bibinfo {author}
  {\bibfnamefont {K.}~\bibnamefont {Jooya}}, \bibinfo {author} {\bibfnamefont
  {P.}~\bibnamefont {Eichler}}, \bibinfo {author} {\bibfnamefont
  {X.}~\bibnamefont {Jiang}}, \emph {et~al.},\ }\href@noop {} {\bibfield
  {journal} {\bibinfo  {journal} {Nature}\ }\textbf {\bibinfo {volume} {604}},\
  \bibinfo {pages} {457} (\bibinfo {year} {2022})}\BibitemShut {NoStop}%
\bibitem [{\citenamefont {Evered}\ \emph {et~al.}(2023)\citenamefont {Evered},
  \citenamefont {Bluvstein}, \citenamefont {Kalinowski}, \citenamefont {Ebadi},
  \citenamefont {Manovitz}, \citenamefont {Zhou}, \citenamefont {Li},
  \citenamefont {Geim} \emph {et~al.}}]{evered2023}%
  \BibitemOpen
  \bibfield  {author} {\bibinfo {author} {\bibfnamefont {S.~J.}\ \bibnamefont
  {Evered}}, \bibinfo {author} {\bibfnamefont {D.}~\bibnamefont {Bluvstein}},
  \bibinfo {author} {\bibfnamefont {M.}~\bibnamefont {Kalinowski}}, \bibinfo
  {author} {\bibfnamefont {S.}~\bibnamefont {Ebadi}}, \bibinfo {author}
  {\bibfnamefont {T.}~\bibnamefont {Manovitz}}, \bibinfo {author}
  {\bibfnamefont {H.}~\bibnamefont {Zhou}}, \bibinfo {author} {\bibfnamefont
  {S.~H.}\ \bibnamefont {Li}}, \bibinfo {author} {\bibfnamefont {A.~A.}\
  \bibnamefont {Geim}}, \emph {et~al.},\ }\href@noop {} {\bibfield  {journal}
  {\bibinfo  {journal} {Nature}\ }\textbf {\bibinfo {volume} {622}},\ \bibinfo
  {pages} {268} (\bibinfo {year} {2023})}\BibitemShut {NoStop}%
\bibitem [{\citenamefont {Bluvstein}\ \emph {et~al.}(2024)\citenamefont
  {Bluvstein}, \citenamefont {Evered}, \citenamefont {Geim}, \citenamefont
  {Li}, \citenamefont {Zhou}, \citenamefont {Manovitz}, \citenamefont {Ebadi},
  \citenamefont {Cain} \emph {et~al.}}]{bluvstein2024}%
  \BibitemOpen
  \bibfield  {author} {\bibinfo {author} {\bibfnamefont {D.}~\bibnamefont
  {Bluvstein}}, \bibinfo {author} {\bibfnamefont {S.~J.}\ \bibnamefont
  {Evered}}, \bibinfo {author} {\bibfnamefont {A.~A.}\ \bibnamefont {Geim}},
  \bibinfo {author} {\bibfnamefont {S.~H.}\ \bibnamefont {Li}}, \bibinfo
  {author} {\bibfnamefont {H.}~\bibnamefont {Zhou}}, \bibinfo {author}
  {\bibfnamefont {T.}~\bibnamefont {Manovitz}}, \bibinfo {author}
  {\bibfnamefont {S.}~\bibnamefont {Ebadi}}, \bibinfo {author} {\bibfnamefont
  {M.}~\bibnamefont {Cain}}, \emph {et~al.},\ }\href@noop {} {\bibfield
  {journal} {\bibinfo  {journal} {Nature}\ }\textbf {\bibinfo {volume} {626}},\
  \bibinfo {pages} {58} (\bibinfo {year} {2024})}\BibitemShut {NoStop}%
\bibitem [{\citenamefont {Reiserer}(2022)}]{reiserer2022}%
  \BibitemOpen
  \bibfield  {author} {\bibinfo {author} {\bibfnamefont {A.}~\bibnamefont
  {Reiserer}},\ }\href@noop {} {\bibfield  {journal} {\bibinfo  {journal} {Rev.
  Mod. Phys.}\ }\textbf {\bibinfo {volume} {94}},\ \bibinfo {pages} {041003}
  (\bibinfo {year} {2022})}\BibitemShut {NoStop}%
\bibitem [{\citenamefont {Covey}\ \emph {et~al.}(2023)\citenamefont {Covey},
  \citenamefont {Weinfurter},\ and\ \citenamefont {Bernien}}]{covey2023}%
  \BibitemOpen
  \bibfield  {author} {\bibinfo {author} {\bibfnamefont {J.~P.}\ \bibnamefont
  {Covey}}, \bibinfo {author} {\bibfnamefont {H.}~\bibnamefont {Weinfurter}},\
  and\ \bibinfo {author} {\bibfnamefont {H.}~\bibnamefont {Bernien}},\
  }\href@noop {} {\bibfield  {journal} {\bibinfo  {journal} {npj Quantum
  Information}\ }\textbf {\bibinfo {volume} {9}},\ \bibinfo {pages} {90}
  (\bibinfo {year} {2023})}\BibitemShut {NoStop}%
\bibitem [{\citenamefont {Sunami}\ \emph {et~al.}(2025)\citenamefont {Sunami},
  \citenamefont {Tamiya}, \citenamefont {Inoue}, \citenamefont {Yamasaki},\
  and\ \citenamefont {Goban}}]{sunami2025}%
  \BibitemOpen
  \bibfield  {author} {\bibinfo {author} {\bibfnamefont {S.}~\bibnamefont
  {Sunami}}, \bibinfo {author} {\bibfnamefont {S.}~\bibnamefont {Tamiya}},
  \bibinfo {author} {\bibfnamefont {R.}~\bibnamefont {Inoue}}, \bibinfo
  {author} {\bibfnamefont {H.}~\bibnamefont {Yamasaki}},\ and\ \bibinfo
  {author} {\bibfnamefont {A.}~\bibnamefont {Goban}},\ }\href@noop {}
  {\bibfield  {journal} {\bibinfo  {journal} {PRX Quantum}\ }\textbf {\bibinfo
  {volume} {6}},\ \bibinfo {pages} {010101} (\bibinfo {year}
  {2025})}\BibitemShut {NoStop}%
\bibitem [{\citenamefont {Kato}\ and\ \citenamefont {Aoki}(2015)}]{kato2015}%
  \BibitemOpen
  \bibfield  {author} {\bibinfo {author} {\bibfnamefont {S.}~\bibnamefont
  {Kato}}\ and\ \bibinfo {author} {\bibfnamefont {T.}~\bibnamefont {Aoki}},\
  }\href@noop {} {\bibfield  {journal} {\bibinfo  {journal} {Phys. Rev. Lett.}\
  }\textbf {\bibinfo {volume} {115}},\ \bibinfo {pages} {093603} (\bibinfo
  {year} {2015})}\BibitemShut {NoStop}%
\bibitem [{\citenamefont {Ruddell}\ \emph {et~al.}(2020)\citenamefont
  {Ruddell}, \citenamefont {Webb}, \citenamefont {Takahata}, \citenamefont
  {Kato},\ and\ \citenamefont {Aoki}}]{ruddell2020}%
  \BibitemOpen
  \bibfield  {author} {\bibinfo {author} {\bibfnamefont {S.~K.}\ \bibnamefont
  {Ruddell}}, \bibinfo {author} {\bibfnamefont {K.~E.}\ \bibnamefont {Webb}},
  \bibinfo {author} {\bibfnamefont {M.}~\bibnamefont {Takahata}}, \bibinfo
  {author} {\bibfnamefont {S.}~\bibnamefont {Kato}},\ and\ \bibinfo {author}
  {\bibfnamefont {T.}~\bibnamefont {Aoki}},\ }\href@noop {} {\bibfield
  {journal} {\bibinfo  {journal} {Opt. Lett.}\ }\textbf {\bibinfo {volume}
  {45}},\ \bibinfo {pages} {4875} (\bibinfo {year} {2020})}\BibitemShut
  {NoStop}%
\bibitem [{\citenamefont {Yimsiriwattana}\ and\ \citenamefont
  {Lomonaco~Jr}(2004)}]{yimsiriwattana2004}%
  \BibitemOpen
  \bibfield  {author} {\bibinfo {author} {\bibfnamefont {A.}~\bibnamefont
  {Yimsiriwattana}}\ and\ \bibinfo {author} {\bibfnamefont {S.~J.}\
  \bibnamefont {Lomonaco~Jr}},\ }in\ \href@noop {} {\emph {\bibinfo {booktitle}
  {Quantum Information and Computation II}}},\ Vol.\ \bibinfo {volume} {5436}\
  (\bibinfo  {publisher} {SPIE},\ \bibinfo {year} {2004})\ pp.\ \bibinfo
  {pages} {360--372}\BibitemShut {NoStop}%
\bibitem [{\citenamefont {Van~Meter}\ \emph {et~al.}(2007)\citenamefont
  {Van~Meter}, \citenamefont {Nemoto},\ and\ \citenamefont
  {Munro}}]{vanmeter2007}%
  \BibitemOpen
  \bibfield  {author} {\bibinfo {author} {\bibfnamefont {R.}~\bibnamefont
  {Van~Meter}}, \bibinfo {author} {\bibfnamefont {K.}~\bibnamefont {Nemoto}},\
  and\ \bibinfo {author} {\bibfnamefont {W.}~\bibnamefont {Munro}},\
  }\href@noop {} {\bibfield  {journal} {\bibinfo  {journal} {IEEE Transactions
  on Computers}\ }\textbf {\bibinfo {volume} {56}},\ \bibinfo {pages} {1643}
  (\bibinfo {year} {2007})}\BibitemShut {NoStop}%
\bibitem [{\citenamefont {Beals}\ \emph {et~al.}(2013)\citenamefont {Beals},
  \citenamefont {Brierley}, \citenamefont {Gray}, \citenamefont {Harrow},
  \citenamefont {Kutin}, \citenamefont {Linden}, \citenamefont {Shepherd},\
  and\ \citenamefont {Stather}}]{beals2013}%
  \BibitemOpen
  \bibfield  {author} {\bibinfo {author} {\bibfnamefont {R.}~\bibnamefont
  {Beals}}, \bibinfo {author} {\bibfnamefont {S.}~\bibnamefont {Brierley}},
  \bibinfo {author} {\bibfnamefont {O.}~\bibnamefont {Gray}}, \bibinfo {author}
  {\bibfnamefont {A.~W.}\ \bibnamefont {Harrow}}, \bibinfo {author}
  {\bibfnamefont {S.}~\bibnamefont {Kutin}}, \bibinfo {author} {\bibfnamefont
  {N.}~\bibnamefont {Linden}}, \bibinfo {author} {\bibfnamefont
  {D.}~\bibnamefont {Shepherd}},\ and\ \bibinfo {author} {\bibfnamefont
  {M.}~\bibnamefont {Stather}},\ }\href@noop {} {\bibfield  {journal} {\bibinfo
   {journal} {Proceedings of the Royal Society A: Mathematical, Physical and
  Engineering Sciences}\ }\textbf {\bibinfo {volume} {469}},\ \bibinfo {pages}
  {20120686} (\bibinfo {year} {2013})}\BibitemShut {NoStop}%
\bibitem [{\citenamefont {Van~Meter}\ and\ \citenamefont
  {Devitt}(2016)}]{vanmeter2016}%
  \BibitemOpen
  \bibfield  {author} {\bibinfo {author} {\bibfnamefont {R.}~\bibnamefont
  {Van~Meter}}\ and\ \bibinfo {author} {\bibfnamefont {S.~J.}\ \bibnamefont
  {Devitt}},\ }\href@noop {} {\bibfield  {journal} {\bibinfo  {journal}
  {Computer}\ }\textbf {\bibinfo {volume} {49}},\ \bibinfo {pages} {31}
  (\bibinfo {year} {2016})}\BibitemShut {NoStop}%
\bibitem [{\citenamefont {Caleffi}\ \emph {et~al.}(2024)\citenamefont
  {Caleffi}, \citenamefont {Amoretti}, \citenamefont {Ferrari}, \citenamefont
  {Illiano}, \citenamefont {Manzalini},\ and\ \citenamefont
  {Cacciapuoti}}]{caleffi2024}%
  \BibitemOpen
  \bibfield  {author} {\bibinfo {author} {\bibfnamefont {M.}~\bibnamefont
  {Caleffi}}, \bibinfo {author} {\bibfnamefont {M.}~\bibnamefont {Amoretti}},
  \bibinfo {author} {\bibfnamefont {D.}~\bibnamefont {Ferrari}}, \bibinfo
  {author} {\bibfnamefont {J.}~\bibnamefont {Illiano}}, \bibinfo {author}
  {\bibfnamefont {A.}~\bibnamefont {Manzalini}},\ and\ \bibinfo {author}
  {\bibfnamefont {A.~S.}\ \bibnamefont {Cacciapuoti}},\ }\href@noop {}
  {\bibfield  {journal} {\bibinfo  {journal} {Computer Networks}\ }\textbf
  {\bibinfo {volume} {254}},\ \bibinfo {pages} {110672} (\bibinfo {year}
  {2024})}\BibitemShut {NoStop}%
\bibitem [{\citenamefont {Cohen}\ \emph {et~al.}(2022)\citenamefont {Cohen},
  \citenamefont {Kim}, \citenamefont {Bartlett},\ and\ \citenamefont
  {Brown}}]{cohen2022}%
  \BibitemOpen
  \bibfield  {author} {\bibinfo {author} {\bibfnamefont {L.~Z.}\ \bibnamefont
  {Cohen}}, \bibinfo {author} {\bibfnamefont {I.~H.}\ \bibnamefont {Kim}},
  \bibinfo {author} {\bibfnamefont {S.~D.}\ \bibnamefont {Bartlett}},\ and\
  \bibinfo {author} {\bibfnamefont {B.~J.}\ \bibnamefont {Brown}},\ }\href@noop
  {} {\bibfield  {journal} {\bibinfo  {journal} {Science Advances}\ }\textbf
  {\bibinfo {volume} {8}},\ \bibinfo {pages} {eabn1717} (\bibinfo {year}
  {2022})}\BibitemShut {NoStop}%
\bibitem [{\citenamefont {Tremblay}\ \emph {et~al.}(2022)\citenamefont
  {Tremblay}, \citenamefont {Delfosse},\ and\ \citenamefont
  {Beverland}}]{tremblay2022}%
  \BibitemOpen
  \bibfield  {author} {\bibinfo {author} {\bibfnamefont {M.~A.}\ \bibnamefont
  {Tremblay}}, \bibinfo {author} {\bibfnamefont {N.}~\bibnamefont {Delfosse}},\
  and\ \bibinfo {author} {\bibfnamefont {M.~E.}\ \bibnamefont {Beverland}},\
  }\href@noop {} {\bibfield  {journal} {\bibinfo  {journal} {Phys. Rev. Lett.}\
  }\textbf {\bibinfo {volume} {129}},\ \bibinfo {pages} {050504} (\bibinfo
  {year} {2022})}\BibitemShut {NoStop}%
\bibitem [{\citenamefont {Yamasaki}\ and\ \citenamefont
  {Koashi}(2024)}]{yamasaki2024}%
  \BibitemOpen
  \bibfield  {author} {\bibinfo {author} {\bibfnamefont {H.}~\bibnamefont
  {Yamasaki}}\ and\ \bibinfo {author} {\bibfnamefont {M.}~\bibnamefont
  {Koashi}},\ }\href@noop {} {\bibfield  {journal} {\bibinfo  {journal} {Nat.
  Phys.}\ }\textbf {\bibinfo {volume} {20}},\ \bibinfo {pages} {247} (\bibinfo
  {year} {2024})}\BibitemShut {NoStop}%
\bibitem [{\citenamefont {Bravyi}\ \emph {et~al.}(2024)\citenamefont {Bravyi},
  \citenamefont {Cross}, \citenamefont {Gambetta}, \citenamefont {Maslov},
  \citenamefont {Rall},\ and\ \citenamefont {Yoder}}]{bravyi2024}%
  \BibitemOpen
  \bibfield  {author} {\bibinfo {author} {\bibfnamefont {S.}~\bibnamefont
  {Bravyi}}, \bibinfo {author} {\bibfnamefont {A.~W.}\ \bibnamefont {Cross}},
  \bibinfo {author} {\bibfnamefont {J.~M.}\ \bibnamefont {Gambetta}}, \bibinfo
  {author} {\bibfnamefont {D.}~\bibnamefont {Maslov}}, \bibinfo {author}
  {\bibfnamefont {P.}~\bibnamefont {Rall}},\ and\ \bibinfo {author}
  {\bibfnamefont {T.~J.}\ \bibnamefont {Yoder}},\ }\href@noop {} {\bibfield
  {journal} {\bibinfo  {journal} {Nature}\ }\textbf {\bibinfo {volume} {627}},\
  \bibinfo {pages} {778} (\bibinfo {year} {2024})}\BibitemShut {NoStop}%
\bibitem [{\citenamefont {Xu}\ \emph {et~al.}(2024)\citenamefont {Xu},
  \citenamefont {Bonilla~Ataides}, \citenamefont {Pattison}, \citenamefont
  {Raveendran}, \citenamefont {Bluvstein}, \citenamefont {Wurtz}, \citenamefont
  {Vasi{\'{c}}}, \citenamefont {Lukin} \emph {et~al.}}]{xu2024}%
  \BibitemOpen
  \bibfield  {author} {\bibinfo {author} {\bibfnamefont {Q.}~\bibnamefont
  {Xu}}, \bibinfo {author} {\bibfnamefont {J.~P.}\ \bibnamefont
  {Bonilla~Ataides}}, \bibinfo {author} {\bibfnamefont {C.~A.}\ \bibnamefont
  {Pattison}}, \bibinfo {author} {\bibfnamefont {N.}~\bibnamefont
  {Raveendran}}, \bibinfo {author} {\bibfnamefont {D.}~\bibnamefont
  {Bluvstein}}, \bibinfo {author} {\bibfnamefont {J.}~\bibnamefont {Wurtz}},
  \bibinfo {author} {\bibfnamefont {B.}~\bibnamefont {Vasi{\'{c}}}}, \bibinfo
  {author} {\bibfnamefont {M.~D.}\ \bibnamefont {Lukin}}, \emph {et~al.},\
  }\href@noop {} {\bibfield  {journal} {\bibinfo  {journal} {Nat. Phys.}\
  }\textbf {\bibinfo {volume} {20}},\ \bibinfo {pages} {1084} (\bibinfo {year}
  {2024})}\BibitemShut {NoStop}%
\bibitem [{\citenamefont {Pecorari}\ \emph {et~al.}(2025)\citenamefont
  {Pecorari}, \citenamefont {Jandura}, \citenamefont {Brennen},\ and\
  \citenamefont {Pupillo}}]{pecorari2025}%
  \BibitemOpen
  \bibfield  {author} {\bibinfo {author} {\bibfnamefont {L.}~\bibnamefont
  {Pecorari}}, \bibinfo {author} {\bibfnamefont {S.}~\bibnamefont {Jandura}},
  \bibinfo {author} {\bibfnamefont {G.~K.}\ \bibnamefont {Brennen}},\ and\
  \bibinfo {author} {\bibfnamefont {G.}~\bibnamefont {Pupillo}},\ }\href@noop
  {} {\bibfield  {journal} {\bibinfo  {journal} {Nat. Commun.}\ }\textbf
  {\bibinfo {volume} {16}},\ \bibinfo {pages} {1111} (\bibinfo {year}
  {2025})}\BibitemShut {NoStop}%
\bibitem [{\citenamefont {Pellizzari}\ \emph {et~al.}(1995)\citenamefont
  {Pellizzari}, \citenamefont {Gardiner}, \citenamefont {Cirac},\ and\
  \citenamefont {Zoller}}]{pellizzari1995}%
  \BibitemOpen
  \bibfield  {author} {\bibinfo {author} {\bibfnamefont {T.}~\bibnamefont
  {Pellizzari}}, \bibinfo {author} {\bibfnamefont {S.~A.}\ \bibnamefont
  {Gardiner}}, \bibinfo {author} {\bibfnamefont {J.~I.}\ \bibnamefont
  {Cirac}},\ and\ \bibinfo {author} {\bibfnamefont {P.}~\bibnamefont
  {Zoller}},\ }\href@noop {} {\bibfield  {journal} {\bibinfo  {journal} {Phys.
  Rev. Lett.}\ }\textbf {\bibinfo {volume} {75}},\ \bibinfo {pages} {3788}
  (\bibinfo {year} {1995})}\BibitemShut {NoStop}%
\bibitem [{\citenamefont {Zheng}\ and\ \citenamefont {Guo}(2000)}]{zheng2000}%
  \BibitemOpen
  \bibfield  {author} {\bibinfo {author} {\bibfnamefont {S.-B.}\ \bibnamefont
  {Zheng}}\ and\ \bibinfo {author} {\bibfnamefont {G.-C.}\ \bibnamefont
  {Guo}},\ }\href@noop {} {\bibfield  {journal} {\bibinfo  {journal} {Phys.
  Rev. Lett.}\ }\textbf {\bibinfo {volume} {85}},\ \bibinfo {pages} {2392}
  (\bibinfo {year} {2000})}\BibitemShut {NoStop}%
\bibitem [{\citenamefont {Duan}\ and\ \citenamefont {Kimble}(2004)}]{duan2004}%
  \BibitemOpen
  \bibfield  {author} {\bibinfo {author} {\bibfnamefont {L.-M.}\ \bibnamefont
  {Duan}}\ and\ \bibinfo {author} {\bibfnamefont {H.~J.}\ \bibnamefont
  {Kimble}},\ }\href@noop {} {\bibfield  {journal} {\bibinfo  {journal} {Phys.
  Rev. Lett.}\ }\textbf {\bibinfo {volume} {92}},\ \bibinfo {pages} {127902}
  (\bibinfo {year} {2004})}\BibitemShut {NoStop}%
\bibitem [{\citenamefont {Xiao}\ \emph {et~al.}(2004)\citenamefont {Xiao},
  \citenamefont {Lin}, \citenamefont {Gao}, \citenamefont {Yang}, \citenamefont
  {Han},\ and\ \citenamefont {Guo}}]{xiao2004}%
  \BibitemOpen
  \bibfield  {author} {\bibinfo {author} {\bibfnamefont {Y.-F.}\ \bibnamefont
  {Xiao}}, \bibinfo {author} {\bibfnamefont {X.-M.}\ \bibnamefont {Lin}},
  \bibinfo {author} {\bibfnamefont {J.}~\bibnamefont {Gao}}, \bibinfo {author}
  {\bibfnamefont {Y.}~\bibnamefont {Yang}}, \bibinfo {author} {\bibfnamefont
  {Z.-F.}\ \bibnamefont {Han}},\ and\ \bibinfo {author} {\bibfnamefont {G.-C.}\
  \bibnamefont {Guo}},\ }\href@noop {} {\bibfield  {journal} {\bibinfo
  {journal} {Phys. Rev. A}\ }\textbf {\bibinfo {volume} {70}},\ \bibinfo
  {pages} {042314} (\bibinfo {year} {2004})}\BibitemShut {NoStop}%
\bibitem [{\citenamefont {Duan}\ \emph {et~al.}(2005)\citenamefont {Duan},
  \citenamefont {Wang},\ and\ \citenamefont {Kimble}}]{duan2005}%
  \BibitemOpen
  \bibfield  {author} {\bibinfo {author} {\bibfnamefont {L.-M.}\ \bibnamefont
  {Duan}}, \bibinfo {author} {\bibfnamefont {B.}~\bibnamefont {Wang}},\ and\
  \bibinfo {author} {\bibfnamefont {H.~J.}\ \bibnamefont {Kimble}},\
  }\href@noop {} {\bibfield  {journal} {\bibinfo  {journal} {Phys. Rev. A}\
  }\textbf {\bibinfo {volume} {72}},\ \bibinfo {pages} {032333} (\bibinfo
  {year} {2005})}\BibitemShut {NoStop}%
\bibitem [{\citenamefont {Koshino}\ \emph {et~al.}(2010)\citenamefont
  {Koshino}, \citenamefont {Ishizaka},\ and\ \citenamefont
  {Nakamura}}]{koshino2010}%
  \BibitemOpen
  \bibfield  {author} {\bibinfo {author} {\bibfnamefont {K.}~\bibnamefont
  {Koshino}}, \bibinfo {author} {\bibfnamefont {S.}~\bibnamefont {Ishizaka}},\
  and\ \bibinfo {author} {\bibfnamefont {Y.}~\bibnamefont {Nakamura}},\
  }\href@noop {} {\bibfield  {journal} {\bibinfo  {journal} {Phys. Rev. A}\
  }\textbf {\bibinfo {volume} {82}},\ \bibinfo {pages} {010301} (\bibinfo
  {year} {2010})}\BibitemShut {NoStop}%
\bibitem [{\citenamefont {Reiserer}\ \emph {et~al.}(2014)\citenamefont
  {Reiserer}, \citenamefont {Kalb}, \citenamefont {Rempe},\ and\ \citenamefont
  {Ritter}}]{reiserer2014}%
  \BibitemOpen
  \bibfield  {author} {\bibinfo {author} {\bibfnamefont {A.}~\bibnamefont
  {Reiserer}}, \bibinfo {author} {\bibfnamefont {N.}~\bibnamefont {Kalb}},
  \bibinfo {author} {\bibfnamefont {G.}~\bibnamefont {Rempe}},\ and\ \bibinfo
  {author} {\bibfnamefont {S.}~\bibnamefont {Ritter}},\ }\href@noop {}
  {\bibfield  {journal} {\bibinfo  {journal} {Nature}\ }\textbf {\bibinfo
  {volume} {508}},\ \bibinfo {pages} {237} (\bibinfo {year}
  {2014})}\BibitemShut {NoStop}%
\bibitem [{\citenamefont {Hacker}\ \emph {et~al.}(2016)\citenamefont {Hacker},
  \citenamefont {Welte}, \citenamefont {Rempe},\ and\ \citenamefont
  {Ritter}}]{hacker2016}%
  \BibitemOpen
  \bibfield  {author} {\bibinfo {author} {\bibfnamefont {B.}~\bibnamefont
  {Hacker}}, \bibinfo {author} {\bibfnamefont {S.}~\bibnamefont {Welte}},
  \bibinfo {author} {\bibfnamefont {G.}~\bibnamefont {Rempe}},\ and\ \bibinfo
  {author} {\bibfnamefont {S.}~\bibnamefont {Ritter}},\ }\href@noop {}
  {\bibfield  {journal} {\bibinfo  {journal} {Nature}\ }\textbf {\bibinfo
  {volume} {536}},\ \bibinfo {pages} {193} (\bibinfo {year}
  {2016})}\BibitemShut {NoStop}%
\bibitem [{\citenamefont {Daiss}\ \emph {et~al.}(2021)\citenamefont {Daiss},
  \citenamefont {Langenfeld}, \citenamefont {Welte}, \citenamefont {Distante},
  \citenamefont {Thomas}, \citenamefont {Hartung}, \citenamefont {Morin},\ and\
  \citenamefont {Rempe}}]{daiss2021}%
  \BibitemOpen
  \bibfield  {author} {\bibinfo {author} {\bibfnamefont {S.}~\bibnamefont
  {Daiss}}, \bibinfo {author} {\bibfnamefont {S.}~\bibnamefont {Langenfeld}},
  \bibinfo {author} {\bibfnamefont {S.}~\bibnamefont {Welte}}, \bibinfo
  {author} {\bibfnamefont {E.}~\bibnamefont {Distante}}, \bibinfo {author}
  {\bibfnamefont {P.}~\bibnamefont {Thomas}}, \bibinfo {author} {\bibfnamefont
  {L.}~\bibnamefont {Hartung}}, \bibinfo {author} {\bibfnamefont
  {O.}~\bibnamefont {Morin}},\ and\ \bibinfo {author} {\bibfnamefont
  {G.}~\bibnamefont {Rempe}},\ }\href@noop {} {\bibfield  {journal} {\bibinfo
  {journal} {Science}\ }\textbf {\bibinfo {volume} {371}},\ \bibinfo {pages}
  {614} (\bibinfo {year} {2021})}\BibitemShut {NoStop}%
\bibitem [{\citenamefont {Goto}\ and\ \citenamefont
  {Ichimura}(2010)}]{goto2010}%
  \BibitemOpen
  \bibfield  {author} {\bibinfo {author} {\bibfnamefont {H.}~\bibnamefont
  {Goto}}\ and\ \bibinfo {author} {\bibfnamefont {K.}~\bibnamefont
  {Ichimura}},\ }\href@noop {} {\bibfield  {journal} {\bibinfo  {journal}
  {Phys. Rev. A}\ }\textbf {\bibinfo {volume} {82}},\ \bibinfo {pages} {032311}
  (\bibinfo {year} {2010})}\BibitemShut {NoStop}%
\bibitem [{\citenamefont {Breuckmann}\ and\ \citenamefont
  {Eberhardt}(2021)}]{breuckmann2021}%
  \BibitemOpen
  \bibfield  {author} {\bibinfo {author} {\bibfnamefont {N.~P.}\ \bibnamefont
  {Breuckmann}}\ and\ \bibinfo {author} {\bibfnamefont {J.~N.}\ \bibnamefont
  {Eberhardt}},\ }\href@noop {} {\bibfield  {journal} {\bibinfo  {journal} {PRX
  Quantum}\ }\textbf {\bibinfo {volume} {2}},\ \bibinfo {pages} {040101}
  (\bibinfo {year} {2021})}\BibitemShut {NoStop}%
\bibitem [{\citenamefont {Asaoka}\ \emph {et~al.}(2021)\citenamefont {Asaoka},
  \citenamefont {Tokunaga}, \citenamefont {Kanamoto}, \citenamefont {Goto},\
  and\ \citenamefont {Aoki}}]{asaoka2021}%
  \BibitemOpen
  \bibfield  {author} {\bibinfo {author} {\bibfnamefont {R.}~\bibnamefont
  {Asaoka}}, \bibinfo {author} {\bibfnamefont {Y.}~\bibnamefont {Tokunaga}},
  \bibinfo {author} {\bibfnamefont {R.}~\bibnamefont {Kanamoto}}, \bibinfo
  {author} {\bibfnamefont {H.}~\bibnamefont {Goto}},\ and\ \bibinfo {author}
  {\bibfnamefont {T.}~\bibnamefont {Aoki}},\ }\href@noop {} {\bibfield
  {journal} {\bibinfo  {journal} {Phys. Rev. A}\ }\textbf {\bibinfo {volume}
  {104}},\ \bibinfo {pages} {043702} (\bibinfo {year} {2021})}\BibitemShut
  {NoStop}%
\bibitem [{\citenamefont {Walls}\ and\ \citenamefont
  {Milburn}(1994)}]{walls1994}%
  \BibitemOpen
  \bibfield  {author} {\bibinfo {author} {\bibfnamefont {D.~F.}\ \bibnamefont
  {Walls}}\ and\ \bibinfo {author} {\bibfnamefont {G.~J.}\ \bibnamefont
  {Milburn}},\ }\href@noop {} {\emph {\bibinfo {title} {Quantum Optics}}}\
  (\bibinfo {address} {Berlin},\ \bibinfo {year} {1994})\BibitemShut {NoStop}%
\bibitem [{\citenamefont {Utsugi}\ \emph {et~al.}(2025)\citenamefont {Utsugi},
  \citenamefont {Asaoka}, \citenamefont {Tokunaga},\ and\ \citenamefont
  {Aoki}}]{utsugi2025}%
  \BibitemOpen
  \bibfield  {author} {\bibinfo {author} {\bibfnamefont {T.}~\bibnamefont
  {Utsugi}}, \bibinfo {author} {\bibfnamefont {R.}~\bibnamefont {Asaoka}},
  \bibinfo {author} {\bibfnamefont {Y.}~\bibnamefont {Tokunaga}},\ and\
  \bibinfo {author} {\bibfnamefont {T.}~\bibnamefont {Aoki}},\ }\href@noop {}
  {\bibfield  {journal} {\bibinfo  {journal} {Phys. Rev. A}\ }\textbf {\bibinfo
  {volume} {111}},\ \bibinfo {pages} {L011701} (\bibinfo {year}
  {2025})}\BibitemShut {NoStop}%
\bibitem [{\citenamefont {Fowler}\ \emph {et~al.}(2012)\citenamefont {Fowler},
  \citenamefont {Mariantoni}, \citenamefont {Martinis},\ and\ \citenamefont
  {Cleland}}]{fowler2012}%
  \BibitemOpen
  \bibfield  {author} {\bibinfo {author} {\bibfnamefont {A.~G.}\ \bibnamefont
  {Fowler}}, \bibinfo {author} {\bibfnamefont {M.}~\bibnamefont {Mariantoni}},
  \bibinfo {author} {\bibfnamefont {J.~M.}\ \bibnamefont {Martinis}},\ and\
  \bibinfo {author} {\bibfnamefont {A.~N.}\ \bibnamefont {Cleland}},\
  }\href@noop {} {\bibfield  {journal} {\bibinfo  {journal} {Physical Review
  A―Atomic, Molecular, and Optical Physics}\ }\textbf {\bibinfo {volume}
  {86}},\ \bibinfo {pages} {032324} (\bibinfo {year} {2012})}\BibitemShut
  {NoStop}%
\bibitem [{\citenamefont {Aaronson}\ and\ \citenamefont
  {Gottesman}(2004)}]{aaronson2004}%
  \BibitemOpen
  \bibfield  {author} {\bibinfo {author} {\bibfnamefont {S.}~\bibnamefont
  {Aaronson}}\ and\ \bibinfo {author} {\bibfnamefont {D.}~\bibnamefont
  {Gottesman}},\ }\href@noop {} {\bibfield  {journal} {\bibinfo  {journal}
  {Phys. Rev. A}\ }\textbf {\bibinfo {volume} {70}},\ \bibinfo {pages} {052328}
  (\bibinfo {year} {2004})}\BibitemShut {NoStop}%
\bibitem [{\citenamefont {Emerson}\ \emph {et~al.}(2007)\citenamefont
  {Emerson}, \citenamefont {Silva}, \citenamefont {Moussa}, \citenamefont
  {Ryan}, \citenamefont {Laforest}, \citenamefont {Baugh}, \citenamefont
  {Cory},\ and\ \citenamefont {Laflamme}}]{emerson2007}%
  \BibitemOpen
  \bibfield  {author} {\bibinfo {author} {\bibfnamefont {J.}~\bibnamefont
  {Emerson}}, \bibinfo {author} {\bibfnamefont {M.}~\bibnamefont {Silva}},
  \bibinfo {author} {\bibfnamefont {O.}~\bibnamefont {Moussa}}, \bibinfo
  {author} {\bibfnamefont {C.}~\bibnamefont {Ryan}}, \bibinfo {author}
  {\bibfnamefont {M.}~\bibnamefont {Laforest}}, \bibinfo {author}
  {\bibfnamefont {J.}~\bibnamefont {Baugh}}, \bibinfo {author} {\bibfnamefont
  {D.~G.}\ \bibnamefont {Cory}},\ and\ \bibinfo {author} {\bibfnamefont
  {R.}~\bibnamefont {Laflamme}},\ }\href@noop {} {\bibfield  {journal}
  {\bibinfo  {journal} {Science}\ }\textbf {\bibinfo {volume} {317}},\ \bibinfo
  {pages} {1893} (\bibinfo {year} {2007})}\BibitemShut {NoStop}%
\bibitem [{\citenamefont {Bendersky}\ \emph {et~al.}(2008)\citenamefont
  {Bendersky}, \citenamefont {Pastawski},\ and\ \citenamefont
  {Paz}}]{bendersky2008}%
  \BibitemOpen
  \bibfield  {author} {\bibinfo {author} {\bibfnamefont {A.}~\bibnamefont
  {Bendersky}}, \bibinfo {author} {\bibfnamefont {F.}~\bibnamefont
  {Pastawski}},\ and\ \bibinfo {author} {\bibfnamefont {J.~P.}\ \bibnamefont
  {Paz}},\ }\href@noop {} {\bibfield  {journal} {\bibinfo  {journal} {Phys.
  Rev. Lett.}\ }\textbf {\bibinfo {volume} {100}},\ \bibinfo {pages} {190403}
  (\bibinfo {year} {2008})}\BibitemShut {NoStop}%
\bibitem [{\citenamefont {Edmonds}(1965)}]{edmonds1965paths}%
  \BibitemOpen
  \bibfield  {author} {\bibinfo {author} {\bibfnamefont {J.}~\bibnamefont
  {Edmonds}},\ }\href@noop {} {\bibfield  {journal} {\bibinfo  {journal}
  {Canadian Journal of mathematics}\ }\textbf {\bibinfo {volume} {17}},\
  \bibinfo {pages} {449} (\bibinfo {year} {1965})}\BibitemShut {NoStop}%
\bibitem [{\citenamefont {Goto}\ \emph {et~al.}(2019)\citenamefont {Goto},
  \citenamefont {Mizukami}, \citenamefont {Tokunaga},\ and\ \citenamefont
  {Aoki}}]{goto2019}%
  \BibitemOpen
  \bibfield  {author} {\bibinfo {author} {\bibfnamefont {H.}~\bibnamefont
  {Goto}}, \bibinfo {author} {\bibfnamefont {S.}~\bibnamefont {Mizukami}},
  \bibinfo {author} {\bibfnamefont {Y.}~\bibnamefont {Tokunaga}},\ and\
  \bibinfo {author} {\bibfnamefont {T.}~\bibnamefont {Aoki}},\ }\href@noop {}
  {\bibfield  {journal} {\bibinfo  {journal} {Physical Review A}\ }\textbf
  {\bibinfo {volume} {99}},\ \bibinfo {pages} {053843} (\bibinfo {year}
  {2019})}\BibitemShut {NoStop}%
\bibitem [{\citenamefont {Schupp}\ \emph {et~al.}(2021)\citenamefont {Schupp},
  \citenamefont {Krcmarsky}, \citenamefont {Krutyanskiy}, \citenamefont
  {Meraner}, \citenamefont {Northup},\ and\ \citenamefont
  {Lanyon}}]{schupp2021}%
  \BibitemOpen
  \bibfield  {author} {\bibinfo {author} {\bibfnamefont {J.}~\bibnamefont
  {Schupp}}, \bibinfo {author} {\bibfnamefont {V.}~\bibnamefont {Krcmarsky}},
  \bibinfo {author} {\bibfnamefont {V.}~\bibnamefont {Krutyanskiy}}, \bibinfo
  {author} {\bibfnamefont {M.}~\bibnamefont {Meraner}}, \bibinfo {author}
  {\bibfnamefont {T.~E.}\ \bibnamefont {Northup}},\ and\ \bibinfo {author}
  {\bibfnamefont {B.~P.}\ \bibnamefont {Lanyon}},\ }\href@noop {} {\bibfield
  {journal} {\bibinfo  {journal} {PRX quantum}\ }\textbf {\bibinfo {volume}
  {2}},\ \bibinfo {pages} {020331} (\bibinfo {year} {2021})}\BibitemShut
  {NoStop}%
\bibitem [{\citenamefont {Gehr}\ \emph {et~al.}(2010)\citenamefont {Gehr},
  \citenamefont {Volz}, \citenamefont {Dubois}, \citenamefont {Steinmetz},
  \citenamefont {Colombe}, \citenamefont {Lev}, \citenamefont {Long},
  \citenamefont {Esteve},\ and\ \citenamefont {Reichel}}]{gehr2010}%
  \BibitemOpen
  \bibfield  {author} {\bibinfo {author} {\bibfnamefont {R.}~\bibnamefont
  {Gehr}}, \bibinfo {author} {\bibfnamefont {J.}~\bibnamefont {Volz}}, \bibinfo
  {author} {\bibfnamefont {G.}~\bibnamefont {Dubois}}, \bibinfo {author}
  {\bibfnamefont {T.}~\bibnamefont {Steinmetz}}, \bibinfo {author}
  {\bibfnamefont {Y.}~\bibnamefont {Colombe}}, \bibinfo {author} {\bibfnamefont
  {B.~L.}\ \bibnamefont {Lev}}, \bibinfo {author} {\bibfnamefont
  {R.}~\bibnamefont {Long}}, \bibinfo {author} {\bibfnamefont {J.}~\bibnamefont
  {Esteve}},\ and\ \bibinfo {author} {\bibfnamefont {J.}~\bibnamefont
  {Reichel}},\ }\href@noop {} {\bibfield  {journal} {\bibinfo  {journal}
  {Physical review letters}\ }\textbf {\bibinfo {volume} {104}},\ \bibinfo
  {pages} {203602} (\bibinfo {year} {2010})}\BibitemShut {NoStop}%
\bibitem [{\citenamefont {Hunger}\ \emph {et~al.}(2010)\citenamefont {Hunger},
  \citenamefont {Steinmetz}, \citenamefont {Colombe}, \citenamefont {Deutsch},
  \citenamefont {H{\"a}nsch},\ and\ \citenamefont {Reichel}}]{hunger2010}%
  \BibitemOpen
  \bibfield  {author} {\bibinfo {author} {\bibfnamefont {D.}~\bibnamefont
  {Hunger}}, \bibinfo {author} {\bibfnamefont {T.}~\bibnamefont {Steinmetz}},
  \bibinfo {author} {\bibfnamefont {Y.}~\bibnamefont {Colombe}}, \bibinfo
  {author} {\bibfnamefont {C.}~\bibnamefont {Deutsch}}, \bibinfo {author}
  {\bibfnamefont {T.~W.}\ \bibnamefont {H{\"a}nsch}},\ and\ \bibinfo {author}
  {\bibfnamefont {J.}~\bibnamefont {Reichel}},\ }\href@noop {} {\bibfield
  {journal} {\bibinfo  {journal} {New Journal of Physics}\ }\textbf {\bibinfo
  {volume} {12}},\ \bibinfo {pages} {065038} (\bibinfo {year}
  {2010})}\BibitemShut {NoStop}%
\bibitem [{\citenamefont {Aoki}\ \emph {et~al.}(2009)\citenamefont {Aoki},
  \citenamefont {Parkins}, \citenamefont {Alton}, \citenamefont {Regal},
  \citenamefont {Dayan}, \citenamefont {Ostby}, \citenamefont {Vahala},\ and\
  \citenamefont {Kimble}}]{aoki2009}%
  \BibitemOpen
  \bibfield  {author} {\bibinfo {author} {\bibfnamefont {T.}~\bibnamefont
  {Aoki}}, \bibinfo {author} {\bibfnamefont {A.}~\bibnamefont {Parkins}},
  \bibinfo {author} {\bibfnamefont {D.}~\bibnamefont {Alton}}, \bibinfo
  {author} {\bibfnamefont {C.}~\bibnamefont {Regal}}, \bibinfo {author}
  {\bibfnamefont {B.}~\bibnamefont {Dayan}}, \bibinfo {author} {\bibfnamefont
  {E.}~\bibnamefont {Ostby}}, \bibinfo {author} {\bibfnamefont {K.~J.}\
  \bibnamefont {Vahala}},\ and\ \bibinfo {author} {\bibfnamefont
  {H.}~\bibnamefont {Kimble}},\ }\href@noop {} {\bibfield  {journal} {\bibinfo
  {journal} {Physical review letters}\ }\textbf {\bibinfo {volume} {102}},\
  \bibinfo {pages} {083601} (\bibinfo {year} {2009})}\BibitemShut {NoStop}%
\bibitem [{\citenamefont {Shomroni}\ \emph {et~al.}(2014)\citenamefont
  {Shomroni}, \citenamefont {Rosenblum}, \citenamefont {Lovsky}, \citenamefont
  {Bechler}, \citenamefont {Guendelman},\ and\ \citenamefont
  {Dayan}}]{shomroni2014}%
  \BibitemOpen
  \bibfield  {author} {\bibinfo {author} {\bibfnamefont {I.}~\bibnamefont
  {Shomroni}}, \bibinfo {author} {\bibfnamefont {S.}~\bibnamefont {Rosenblum}},
  \bibinfo {author} {\bibfnamefont {Y.}~\bibnamefont {Lovsky}}, \bibinfo
  {author} {\bibfnamefont {O.}~\bibnamefont {Bechler}}, \bibinfo {author}
  {\bibfnamefont {G.}~\bibnamefont {Guendelman}},\ and\ \bibinfo {author}
  {\bibfnamefont {B.}~\bibnamefont {Dayan}},\ }\href@noop {} {\bibfield
  {journal} {\bibinfo  {journal} {Science}\ }\textbf {\bibinfo {volume}
  {345}},\ \bibinfo {pages} {903} (\bibinfo {year} {2014})}\BibitemShut
  {NoStop}%
\bibitem [{\citenamefont {Scheucher}\ \emph {et~al.}(2016)\citenamefont
  {Scheucher}, \citenamefont {Hilico}, \citenamefont {Will}, \citenamefont
  {Volz},\ and\ \citenamefont {Rauschenbeutel}}]{scheucher2016}%
  \BibitemOpen
  \bibfield  {author} {\bibinfo {author} {\bibfnamefont {M.}~\bibnamefont
  {Scheucher}}, \bibinfo {author} {\bibfnamefont {A.}~\bibnamefont {Hilico}},
  \bibinfo {author} {\bibfnamefont {E.}~\bibnamefont {Will}}, \bibinfo {author}
  {\bibfnamefont {J.}~\bibnamefont {Volz}},\ and\ \bibinfo {author}
  {\bibfnamefont {A.}~\bibnamefont {Rauschenbeutel}},\ }\href@noop {}
  {\bibfield  {journal} {\bibinfo  {journal} {Science}\ }\textbf {\bibinfo
  {volume} {354}},\ \bibinfo {pages} {1577} (\bibinfo {year}
  {2016})}\BibitemShut {NoStop}%
\bibitem [{\citenamefont {Papon}\ \emph {et~al.}(2019)\citenamefont {Papon},
  \citenamefont {Zhou}, \citenamefont {Thyrrestrup}, \citenamefont {Liu},
  \citenamefont {Stobbe}, \citenamefont {Schott}, \citenamefont {Wieck},
  \citenamefont {Ludwig}, \citenamefont {Lodahl},\ and\ \citenamefont
  {Midolo}}]{papon2019}%
  \BibitemOpen
  \bibfield  {author} {\bibinfo {author} {\bibfnamefont {C.}~\bibnamefont
  {Papon}}, \bibinfo {author} {\bibfnamefont {X.}~\bibnamefont {Zhou}},
  \bibinfo {author} {\bibfnamefont {H.}~\bibnamefont {Thyrrestrup}}, \bibinfo
  {author} {\bibfnamefont {Z.}~\bibnamefont {Liu}}, \bibinfo {author}
  {\bibfnamefont {S.}~\bibnamefont {Stobbe}}, \bibinfo {author} {\bibfnamefont
  {R.}~\bibnamefont {Schott}}, \bibinfo {author} {\bibfnamefont {A.~D.}\
  \bibnamefont {Wieck}}, \bibinfo {author} {\bibfnamefont {A.}~\bibnamefont
  {Ludwig}}, \bibinfo {author} {\bibfnamefont {P.}~\bibnamefont {Lodahl}},\
  and\ \bibinfo {author} {\bibfnamefont {L.}~\bibnamefont {Midolo}},\
  }\href@noop {} {\bibfield  {journal} {\bibinfo  {journal} {Optica}\ }\textbf
  {\bibinfo {volume} {6}},\ \bibinfo {pages} {524} (\bibinfo {year}
  {2019})}\BibitemShut {NoStop}%
\bibitem [{\citenamefont {Ren}\ \emph {et~al.}(2022)\citenamefont {Ren},
  \citenamefont {Ma}, \citenamefont {Xie}, \citenamefont {Li}, \citenamefont
  {Cao},\ and\ \citenamefont {Li}}]{ren2022}%
  \BibitemOpen
  \bibfield  {author} {\bibinfo {author} {\bibfnamefont {Y.-l.}\ \bibnamefont
  {Ren}}, \bibinfo {author} {\bibfnamefont {S.-l.}\ \bibnamefont {Ma}},
  \bibinfo {author} {\bibfnamefont {J.-k.}\ \bibnamefont {Xie}}, \bibinfo
  {author} {\bibfnamefont {X.-k.}\ \bibnamefont {Li}}, \bibinfo {author}
  {\bibfnamefont {M.-t.}\ \bibnamefont {Cao}},\ and\ \bibinfo {author}
  {\bibfnamefont {F.-l.}\ \bibnamefont {Li}},\ }\href@noop {} {\bibfield
  {journal} {\bibinfo  {journal} {Physical Review A}\ }\textbf {\bibinfo
  {volume} {105}},\ \bibinfo {pages} {013711} (\bibinfo {year}
  {2022})}\BibitemShut {NoStop}%
\bibitem [{\citenamefont {Aliferis}\ and\ \citenamefont
  {Preskill}(2008)}]{aliferis2008}%
  \BibitemOpen
  \bibfield  {author} {\bibinfo {author} {\bibfnamefont {P.}~\bibnamefont
  {Aliferis}}\ and\ \bibinfo {author} {\bibfnamefont {J.}~\bibnamefont
  {Preskill}},\ }\href@noop {} {\bibfield  {journal} {\bibinfo  {journal}
  {Physical Review A―Atomic, Molecular, and Optical Physics}\ }\textbf
  {\bibinfo {volume} {78}},\ \bibinfo {pages} {052331} (\bibinfo {year}
  {2008})}\BibitemShut {NoStop}%
\bibitem [{\citenamefont {Tuckett}\ \emph {et~al.}(2018)\citenamefont
  {Tuckett}, \citenamefont {Bartlett},\ and\ \citenamefont
  {Flammia}}]{tuckett2018}%
  \BibitemOpen
  \bibfield  {author} {\bibinfo {author} {\bibfnamefont {D.~K.}\ \bibnamefont
  {Tuckett}}, \bibinfo {author} {\bibfnamefont {S.~D.}\ \bibnamefont
  {Bartlett}},\ and\ \bibinfo {author} {\bibfnamefont {S.~T.}\ \bibnamefont
  {Flammia}},\ }\href@noop {} {\bibfield  {journal} {\bibinfo  {journal}
  {Physical review letters}\ }\textbf {\bibinfo {volume} {120}},\ \bibinfo
  {pages} {050505} (\bibinfo {year} {2018})}\BibitemShut {NoStop}%
\bibitem [{\citenamefont {Tuckett}\ \emph {et~al.}(2020)\citenamefont
  {Tuckett}, \citenamefont {Bartlett}, \citenamefont {Flammia},\ and\
  \citenamefont {Brown}}]{tuckett2020}%
  \BibitemOpen
  \bibfield  {author} {\bibinfo {author} {\bibfnamefont {D.~K.}\ \bibnamefont
  {Tuckett}}, \bibinfo {author} {\bibfnamefont {S.~D.}\ \bibnamefont
  {Bartlett}}, \bibinfo {author} {\bibfnamefont {S.~T.}\ \bibnamefont
  {Flammia}},\ and\ \bibinfo {author} {\bibfnamefont {B.~J.}\ \bibnamefont
  {Brown}},\ }\href@noop {} {\bibfield  {journal} {\bibinfo  {journal}
  {Physical review letters}\ }\textbf {\bibinfo {volume} {124}},\ \bibinfo
  {pages} {130501} (\bibinfo {year} {2020})}\BibitemShut {NoStop}%
\bibitem [{\citenamefont {Darmawan}\ \emph {et~al.}(2021)\citenamefont
  {Darmawan}, \citenamefont {Brown}, \citenamefont {Grimsmo}, \citenamefont
  {Tuckett},\ and\ \citenamefont {Puri}}]{darmawan2021}%
  \BibitemOpen
  \bibfield  {author} {\bibinfo {author} {\bibfnamefont {A.~S.}\ \bibnamefont
  {Darmawan}}, \bibinfo {author} {\bibfnamefont {B.~J.}\ \bibnamefont {Brown}},
  \bibinfo {author} {\bibfnamefont {A.~L.}\ \bibnamefont {Grimsmo}}, \bibinfo
  {author} {\bibfnamefont {D.~K.}\ \bibnamefont {Tuckett}},\ and\ \bibinfo
  {author} {\bibfnamefont {S.}~\bibnamefont {Puri}},\ }\href@noop {} {\bibfield
   {journal} {\bibinfo  {journal} {PRX Quantum}\ }\textbf {\bibinfo {volume}
  {2}},\ \bibinfo {pages} {030345} (\bibinfo {year} {2021})}\BibitemShut
  {NoStop}%
\bibitem [{\citenamefont {Wu}\ \emph {et~al.}(2022)\citenamefont {Wu},
  \citenamefont {Kolkowitz}, \citenamefont {Puri},\ and\ \citenamefont
  {Thompson}}]{wu2022}%
  \BibitemOpen
  \bibfield  {author} {\bibinfo {author} {\bibfnamefont {Y.}~\bibnamefont
  {Wu}}, \bibinfo {author} {\bibfnamefont {S.}~\bibnamefont {Kolkowitz}},
  \bibinfo {author} {\bibfnamefont {S.}~\bibnamefont {Puri}},\ and\ \bibinfo
  {author} {\bibfnamefont {J.~D.}\ \bibnamefont {Thompson}},\ }\href@noop {}
  {\bibfield  {journal} {\bibinfo  {journal} {Nature communications}\ }\textbf
  {\bibinfo {volume} {13}},\ \bibinfo {pages} {4657} (\bibinfo {year}
  {2022})}\BibitemShut {NoStop}%
\bibitem [{\citenamefont {Claes}\ \emph {et~al.}(2023)\citenamefont {Claes},
  \citenamefont {Bourassa},\ and\ \citenamefont {Puri}}]{claes2023}%
  \BibitemOpen
  \bibfield  {author} {\bibinfo {author} {\bibfnamefont {J.}~\bibnamefont
  {Claes}}, \bibinfo {author} {\bibfnamefont {J.~E.}\ \bibnamefont
  {Bourassa}},\ and\ \bibinfo {author} {\bibfnamefont {S.}~\bibnamefont
  {Puri}},\ }\href@noop {} {\bibfield  {journal} {\bibinfo  {journal} {npj
  Quantum Information}\ }\textbf {\bibinfo {volume} {9}},\ \bibinfo {pages} {9}
  (\bibinfo {year} {2023})}\BibitemShut {NoStop}%
\bibitem [{\citenamefont {Sahay}\ \emph {et~al.}(2023)\citenamefont {Sahay},
  \citenamefont {Jin}, \citenamefont {Claes}, \citenamefont {Thompson},\ and\
  \citenamefont {Puri}}]{sahay2023}%
  \BibitemOpen
  \bibfield  {author} {\bibinfo {author} {\bibfnamefont {K.}~\bibnamefont
  {Sahay}}, \bibinfo {author} {\bibfnamefont {J.}~\bibnamefont {Jin}}, \bibinfo
  {author} {\bibfnamefont {J.}~\bibnamefont {Claes}}, \bibinfo {author}
  {\bibfnamefont {J.~D.}\ \bibnamefont {Thompson}},\ and\ \bibinfo {author}
  {\bibfnamefont {S.}~\bibnamefont {Puri}},\ }\href@noop {} {\bibfield
  {journal} {\bibinfo  {journal} {Physical Review X}\ }\textbf {\bibinfo
  {volume} {13}},\ \bibinfo {pages} {041013} (\bibinfo {year}
  {2023})}\BibitemShut {NoStop}%
\bibitem [{\citenamefont {Duckering}\ \emph {et~al.}(2020)\citenamefont
  {Duckering}, \citenamefont {Baker}, \citenamefont {Schuster},\ and\
  \citenamefont {Chong}}]{duckering2020}%
  \BibitemOpen
  \bibfield  {author} {\bibinfo {author} {\bibfnamefont {C.}~\bibnamefont
  {Duckering}}, \bibinfo {author} {\bibfnamefont {J.~M.}\ \bibnamefont
  {Baker}}, \bibinfo {author} {\bibfnamefont {D.~I.}\ \bibnamefont
  {Schuster}},\ and\ \bibinfo {author} {\bibfnamefont {F.~T.}\ \bibnamefont
  {Chong}},\ }\href@noop {} {\bibinfo {title} {Virtualized logical qubits: A
  2.5 d architecture for error-corrected quantum computing}} (\bibinfo {year}
  {2020})\BibitemShut {NoStop}%
\bibitem [{\citenamefont {Ramette}\ \emph {et~al.}(2022)\citenamefont
  {Ramette}, \citenamefont {Sinclair}, \citenamefont {Vendeiro}, \citenamefont
  {Rudelis}, \citenamefont {Cetina},\ and\ \citenamefont
  {Vuleti{\'c}}}]{ramette2022}%
  \BibitemOpen
  \bibfield  {author} {\bibinfo {author} {\bibfnamefont {J.}~\bibnamefont
  {Ramette}}, \bibinfo {author} {\bibfnamefont {J.}~\bibnamefont {Sinclair}},
  \bibinfo {author} {\bibfnamefont {Z.}~\bibnamefont {Vendeiro}}, \bibinfo
  {author} {\bibfnamefont {A.}~\bibnamefont {Rudelis}}, \bibinfo {author}
  {\bibfnamefont {M.}~\bibnamefont {Cetina}},\ and\ \bibinfo {author}
  {\bibfnamefont {V.}~\bibnamefont {Vuleti{\'c}}},\ }\href@noop {} {\bibfield
  {journal} {\bibinfo  {journal} {PRX Quantum}\ }\textbf {\bibinfo {volume}
  {3}},\ \bibinfo {pages} {010344} (\bibinfo {year} {2022})}\BibitemShut
  {NoStop}%
\end{thebibliography}%

\end{document}